\documentclass[fleqn,usenatbib]{mnras}

\usepackage{newtxtext,newtxmath}

\usepackage[T1]{fontenc}
\usepackage[utf8]{inputenc}

\DeclareRobustCommand{\VAN}[3]{#2}
\let\VANthebibliography\thebibliography
\def\thebibliography{\DeclareRobustCommand{\VAN}[3]{##3}\VANthebibliography}

\usepackage{graphicx}
\usepackage{subfig}
\usepackage{amsmath}
\usepackage{amssymb}
\usepackage{tabularx}
\newcommand{\ud}{\mathrm{d}}
\newcommand{\Red}[1]{{\color{black} #1}}

\newcommand{\bm}[1]{\textbf{\color{green} #1}}

\title[Simulations of Magnetised Non-rotating Supernovae]{3D Simulations of Strongly Magnetised Non-Rotating  Supernovae: Explosion Dynamics and Remnant Properties}

\author[Varma et al.]{
Vishnu Varma$^{1,2}$\thanks{E-mail: 
vishnu.rvejayan@monash.edu},
Bernhard M\"uller$^{1}$\thanks{E-mail: 
bernhard.mueller@monash.edu
},
Fabian R.~N. Schneider$^{3,4}$
\\
$^{1}$
School of Physics and Astronomy, 10 College Walk, Monash University, Clayton, VIC 3800, Australia\\
$^{2}$
{Astrophysics Group, Lennard-Jones Laboratories, Keele University, Keele ST5 5BG, UK}\\
$^{3}$
Heidelberger Institut f\"ur Theoretische Studien, Schlo{\ss}-Wolfsbrunnenweg 35, 69118 Heidelberg, Germany\\
$^{4}$
Astronomisches Rechen-Institut, Zentrum f\"ur Astronomie der Universit\"at Heidelberg,
M\"onchhofstraße 12--14, 69120 Heidelberg, Germany
}

\date{Accepted XXX. Received YYY; in original form ZZZ}

\pubyear{2022}

\begin{document}
\label{firstpage}
\pagerange{\pageref{firstpage}--\pageref{lastpage}}
\maketitle

\begin{abstract}
We investigate the impact of strong initial  magnetic fields in core-collapse supernovae of non-rotating progenitors by simulating the collapse and explosion of a $16.9\, M_\odot$ star for a strong- and weak-field case assuming a twisted-torus field with initial central field strengths of $\mathord{\approx}10^{12}\, \mathrm{G}$ and $\mathord{\approx}10^{6}\, \mathrm{G}$. The strong-field model has been set up with a view to the fossil-field scenario for magnetar formation and emulates a pre-collapse field configuration that may occur in massive stars formed by a merger. This model undergoes shock revival already \Red{$100\,\mathrm{ms}$} after bounce and reaches an explosion energy of $9.3\times 10^{50}\,\mathrm{erg}$ at $310\,\mathrm{ms}$, in contrast to a more delayed and less energetic explosion in the weak-field model. The strong magnetic fields help trigger a neutrino-driven explosion early on, which results in a rapid rise and saturation of the explosion energy. Dynamically, the strong initial field leads to a fast build-up of magnetic fields in the gain region to 40\% of kinetic equipartition and also creates sizable pre-shock ram pressure perturbations that are known to be conducive to asymmetric shock expansion. For the strong-field model, we find an extrapolated neutron star kick of $\mathord{\approx}350\,\mathrm{km}\,\mathrm{s}^{-1}$, a spin period of $\mathord{\approx}70\, \mathrm{ms}$, and no spin-kick alignment. The dipole field strength of the proto-neutron star is $2\times 10^{14}\,\mathrm{G}$ by the end of the simulation with a declining trend. Surprisingly, the surface dipole field in the weak-field model is stronger, which argues against a straightforward connection between pre-collapse fields and the birth magnetic fields of neutron stars.
\end{abstract}


\section{Introduction}
\label{sec:intro} 

In the last few decades, there has been significant progress in our understanding of the role of magnetic fields in core-collapse supernovae. This is owed, in part, to rapid improvements in numerical modelling and computing capabilities. 

Since traditional stellar evolution models \citep{Heger2005} have predicted weak pre-collapse magnetic field strengths, most of the progress in understanding magnetic effects in core-collapse supernovae have been focused on the so-called magnetorotational mechanism \citep{Leblanc1970, Bisnovatyi-Kogan1976, Mueller1979, Burrows2007,Obergaulinger2009, Sawai2013, Masada2015, Moesta2015, Bugli2020, Obergaulinger2021}. This mechanism relies on a rapidly differentially rotating star to amplify the weak magnetic fields, e.g., by the magnetorotational instability (MRI; \citealt{Balbus1991, Akiyama2003})
or an $\alpha$-$\Omega$-dynamo in the proto-neutron
star \citep{duncan_92,Thompson1993,Raynaud2020}. Strong magnetic stresses can then drive the explosion, usually in the form of collimated jets. 

However, stellar evolution models also show magnetic braking causes the majority of supernova progenitor cores to rotate slowly \citep{Heger2005, Meynet2011} as the magnetic torques transport angular momentum to the envelope very efficiently. 
Even without magnetic braking, the winds of Galactic stars will prevent rapid rotation at core collapse. These predictions are
bolstered by the white dwarf spin period distribution, which can only be explained if similarly efficient angular momentum transport mechanisms like magnetic torques operate in low-mass stars \citep{suijs_08}.
It is thus expected that the magnetorotational mechanism likely only operates in rare hyperenergetic supernovae explosions
(known as ``hypernovae''), which are sometimes linked to long-duration gamma-ray bursts (GRBs) as well \citep{Woosley2006b}. These scenarios require special evolutionary pathways to limit the spin-down of the progenitor cores
\citep[e.g.,][]{petrovic_05,Woosley2006,georgy_07,georgy_09,cantiello_07,aguilera_20}.

The role of magnetic fields in supernovae outside the magnetorotational mechanism has so far received less attention. Without strong differential rotation, there is no obvious avenue for generating dynamically relevant fields that could become the main driver of an explosion.
Nonetheless, some studies have explored the generation and impact of magnetic fields in non-rotating core-collapse supernovae in recent years.
Early studies by \citet{Endeve2010, Endeve2012} looked to understand the possible role of standing accretion shock instabilities (SASI) to amplify magnetic fields after core-collapse. They found that while the SASI oscillations can amplify the magnetic fields on the neutron star to magnetar strengths, this does not affect global shock dynamics.
Global studies in axisymmetric (2D) simulations by \citet{Obergaulinger2014}, \citet{Matsumoto2020}
and \citet{jardine_22} with MHD and neutrino transport, find that even very strong magnetic fields ($\mathord{\approx} 10^{12}\, \mathrm{G}$), without rotation, only play an auxiliary in neutrino-driven explosions.  However, recently, \citet{MullerVarma2020} have shown that in 3D, even very weak initial magnetic fields as low as $10^{6}\,\mathrm{G}$ can play a subsidiary role in aiding the neutrino-driven mechanism thanks to efficient amplification by a turbulent dynamo. A 3D study by \citet{Matsumoto2022} also showed that in more strongly magnetised models, the amplified magnetic field on large hot bubbles just behind a stalled shock can lead to a faster and more energetic explosion.

Another possibility for dynamically relevant magnetic fields, is that they might already be present in the progenitor. This is known as the fossil-field hypothesis for the generation of magnetic fields in
neutron stars and, in particular, in  magnetars. If the initial fields are strong enough, the compression via core-collapse alone would be sufficient to  amplify the magnetic fields to magnetar strengths of $10^{14\texttt{-}15}\,\mathrm{G}$, through magnetic flux conservation \citep{Braithwaite2004,Spruit2009, ferrario_18}.

To explain magnetars by the fossil-field hypothesis, a remarkably strong field already would need to be present in the progenitor core in a subset of supernova progenitors. Based on a rough coincidence of the magnetar fraction and the fraction of 
roughly $7\%$ of massive stars with unusually strong surface fields \citep{Donati2009, Grunhut2017, Scholler2017}, a connection is often made between strongly magnetised massive stars on the main sequence and magnetars with the underlying assumption that the interior field is likewise strong and not destroyed through the subsequent evolutionary stages up to iron core collapse \citep[see, e.g.,][for the evolution of magnetised, low-mass stars up to white-dwarf formation]{Quentin2018}. Connecting the origin of magnetar fields back to early evolutionary stages is appealing because
stellar mergers might provide one natural avenue for producing a population of strongly magnetised
stars roughly at the rate required to explain magnetars
\citep{SchneiderNature}; other channels for forming strongly magnetised OB stars are also conceivable \citep[see, e.g.,][]{Mestel1999, Mestel2001, Moss2001, Donati2009}, and this work should be viewed as being agnostic about the exact origin of the strong magnetic fields.

Rate arguments in favour of the fossil-field hypothesis are not uncontested, however.
Some population synthesis models indicate that the simple fossil-field hypothesis based on the distribution of surface magnetic field strengths among massive stars cannot simultaneously reproduce both pulsar and magnetar populations \citep{ferrario_06,Makarenko2021}.
It has also been found that the magnetar birth rate is comparable to that of neutron stars (and could be as high as the neutron star birth rate), which may leave too few known stars with surface $B\,>\,10\,\mathrm{kG}$ to account for the number of magnetars \citep{woods_08, keane_08} and raises doubts whether fossil fields are the only means for generating magnetar-strength fields.

With many open questions about the origin of neutron star magnetic fields and recent theoretical results that suggest a broader role of magnetohydrodynamic effects in supernova explosions, further research into the impact of magnetic fields on supernova explosion dynamics and compact remnant birth properties is called for. In this study, we present new simulations of a strongly magnetised, non-rotating neutrino-driven supernova in 3D and compare this to a weakly magnetised model to better understand how the strength of the magnetic field aids the explosion mechanism even without any initial rotation. While 1D stellar evolution models indicate only weak magnetic fields developing in supernova progenitors, our decision to explore the strongly magnetised regime is motivated by
the fossil-field hypothesis and also by the 3D shell burning simulation of \citet{VarmaMuller2021}, which suggests the shell convection can generate rather strong magnetic fields of $10^{10\texttt{-}11}\,\mathrm{G}$ in the oxygen shell.

Our paper is structured as follows: In Section~\ref{sec:initial}, we describe the progenitor model as well as the initial conditions implemented in our simulations, this is followed by a description of the numerical methods of the code we used, \textsc{CoCoNuT-FMT} in Section~\ref{sec:methods}. The results of the simulations are  presented in Section~\ref{sec:results} where we first look at the large-scale explosion dynamics, and analyse the role of the magnetic fields in the explosion, followed by an investigation of the resulting proto-neutron star (PNS) properties. We summarise our results and discuss their implications in Section~\ref{sec:conclusion}.

\section{Progenitor Model and Initial Conditions}
\label{sec:initial} 

We simulate the core collapse of a strongly magnetised, slowly-rotating massive star of $\mathord{16.9\, \mathrm{M_\odot}}$ that is the result of a binary merger of $8\, \mathrm{M_\odot}$ and $9\, \mathrm{M_\odot}$ stars as modelled in \citet{SchneiderNature, Schneider2020} in an attempt to \Red{reproduce} the magnetic star $\tau$~Sco. The merger was simulated using the \textsc{AREPO} code, followed by 1D stellar evolution of the merger product in the \textsc{MESA} code. We map this \textsc{MESA} progenitor model to the \textsc{CoCoNuT-FMT} supernova code at the onset of core collapse when the maximum infall velocity has reached
$1000\, \mathrm{km}\, \mathrm{s}^{-1}$. \Red{With a maximum angular velocity of $2.2\times10^{-8}\, \mathrm{rad}\, \mathrm{s}^{-1}$ the rotation is, in fact, so slow that it can be neglected
altogether, hence we do not map the angular velocity from the \textsc{MESA} progenitor model to \textsc{CoCoNuT-FMT}, and instead simulate the non-rotating case.}

\begin{figure}
   \centering
  \includegraphics[width=\linewidth]{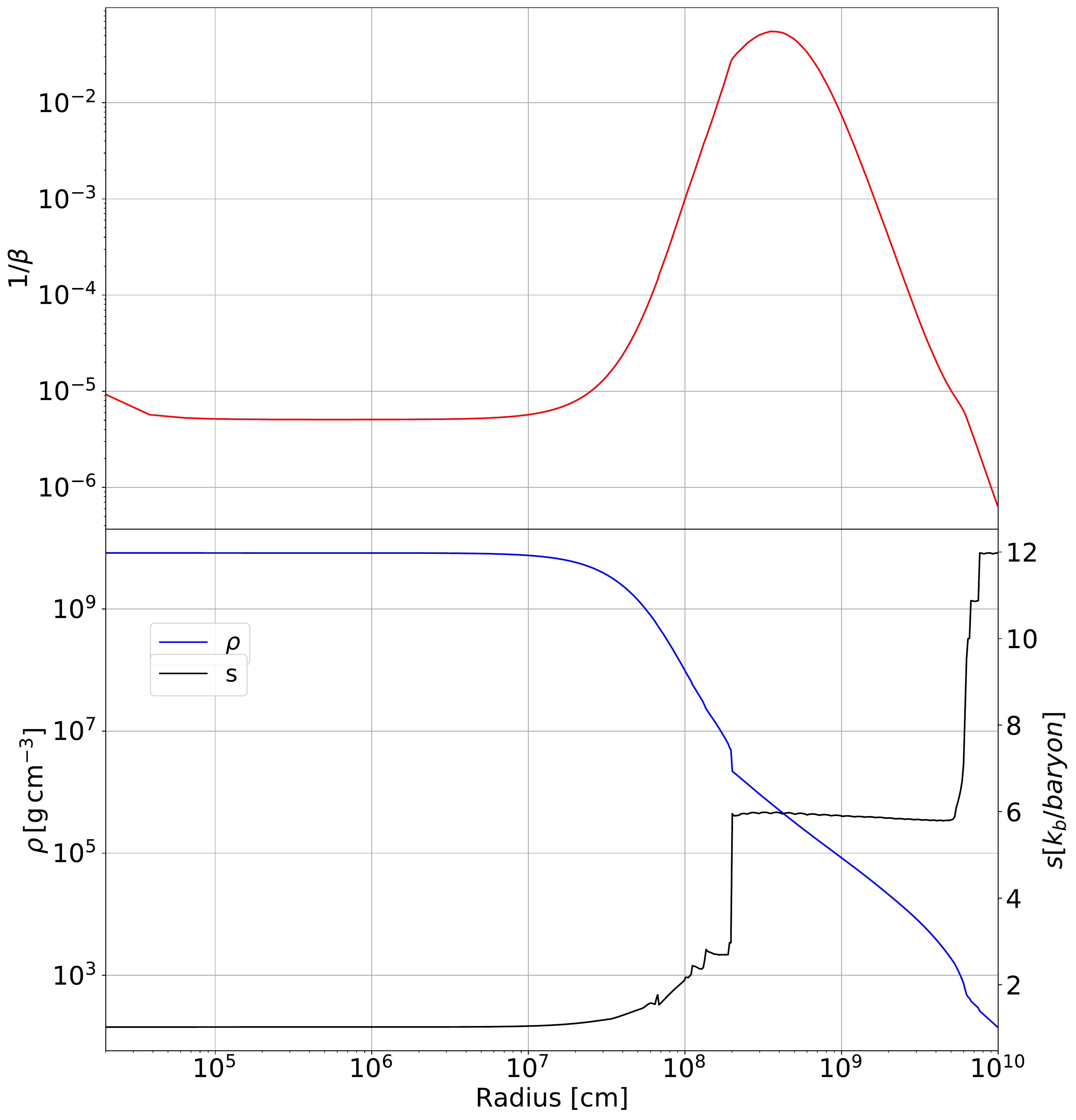}
  \caption{Inverse plasma-$\beta$ at the onset of collapse for the strong-field model (top), and initial density and entropy profiles (bottom) for the progenitor. 
  }
  \label{fig:init}
\end{figure}

The \textsc{MESA} model is evolved until core collapse with the same setup as in \citet{Schneider2021}. The entropy and density profiles mapped are shown in Figure \ref{fig:init}.

Three-dimensional progenitor models of stars with strong fossil fields in the core are presently unavailable and unrealistic because they would require multi-dimensional stellar evolution calculations over secular time scales. We therefore manually specify a field configuration that is compatible with the expectations for a strong fossil field. Prior to mapping the merger progenitor into \textsc{MESA}, the magnetic field geometry of the star from \citet{SchneiderNature} is $80\texttt{-}85\%$ toroidal, but with a sizable poloidal component. This is consistent with the stable twisted-torus structure found in radiative zones found by \citet{Braithwaite2006}.

We hence initialise the magnetic field geometry using an analytical prescription from \citet{Kamchatnov1982} for such twisted-torus fields. Although \citet{Kamchatnov1982} did not prove the stability of this configuration, this is not essential for our purpose. The fields have no time to adjust before core collapse after the beginning of the simulation, and during the collapse even an initially stable magnetic field would become unstable anyway. The key point is that the configuration exhibits substantial magnetic helicity as one would expect from any realistic strong fossil field. 

In spherical polar coordinates, the components $A_r$,
$A_\theta$, and $A_\varphi$ of the vector potential for the initial field configuration are given by
\begin{eqnarray}
    A_r = B_0\frac{(1 + \mathcal{R}^2)\cos\theta}{4(1 + 2\mathcal{R}^2 + \mathcal{R}^4)},
    \label{eq:b_ini_1}
    \\
    A_{\theta} = B_0\frac{(-1 + \mathcal{R}^2)\sin\theta}{4(1 + 2\mathcal{R}^2 + \mathcal{R}^4)},
    \label{eq:b_ini_2}
    \\
    A_{\phi} = B_0\frac{\mathcal{R} \sin\theta}{2(1 + 2\mathcal{R}^2 + \mathcal{R}^4)},
    \label{eq:b_ini_3}
\end{eqnarray}
where $B_0$ is the central (maximum) magnetic field strength, and $\mathcal{R}=r/R_0$ is a rescaled radial coordinate.

As we are investigating the role of magnetic fields in aiding the explosion mechanism for non-rotating supernova progenitors, we run two simulations using the same progenitor with different initial magnetic field strengths, namely a weak-field model, with  $B_0=10^{6}\,\mathrm{G}$ and a strong-field model with $B_0=10^{12}\,\mathrm{G}$ and $R_0=5\times10^8\,\mathrm{cm}$. 

The strong-field model serves as the representative case for the fossil-field magnetar progenitor scenario, whereas the weak-field model provides a control case. Since the magnetic field prescription 
from Equations~(\ref{eq:b_ini_1}--\ref{eq:b_ini_3}) leads to a very sharp drop-off in magnetic field strength with radius, we scale the inner core to have magnetic field strengths of $B_0$, such that for the strong magnetic field case, the pre-supernova oxygen shell has magnetic field strengths of $B=10^{10} - 10^{11}G$, (similar to \citealt{VarmaMuller2021}).
This is similar to the \textsc{MESA} progenitor model which should have a core field strength of $\mathord{\gtrsim} 10^{11} \,\mathrm{G}$, given magnetic flux is conserved from the original stellar merger calculation.

Without 3D progenitor models close to magnetohydrostatic equilibrium, a delicate balance must be struck in the choice of the initial field strength. The initial fields must not be too strong because this would render the initial pre-collapse structure highly inconsistent, but to realistically mimic a strong fossil field that would not have been overwhelmed by turbulent convective motions during the late burning stages, the magnetic stresses ought to be larger than the expected turbulent Reynolds stresses from convection. The configuration chosen in this study places the ratio of magnetic to thermal pressure (inverse plasma-$\beta$) in the oxygen shell at 0.05 as a reasonable compromise (see Figure \ref{fig:init}).

\section{Numerical Methods}
\label{sec:methods} 
For our 3D simulations we employ the Newtonian  magnetohydrodynamic (MHD) version of the  \textsc{CoCoNuT-FMT} code as described in detail in \citet{MullerVarma2020, VarmaMuller2021, Varma2021}.
The MHD equations are solved in spherical
polar coordinates using the HLLC (Harten-Lax-van Leer-Contact) Riemann solver \citep{Gurski2004, Miyoshi2005} and piecewise parabolic reconstruction
\citep{Colella1984}. The divergence-free condition $\nabla\cdot\mathbf{B} = 0$ is maintained using a modified version
of the original hyperbolic divergence cleaning scheme of \citet{Dedner2002} and follows ideas of \citet{Tricco2016} to maintain energy conservation. Compared to the original cleaning method, we rescale
the Lagrange multiplier $\psi$ to
$\hat{\psi}=\psi/c_\mathrm{h}$, where $c_\mathrm{h}$ is the hyperbolic cleaning speed. Details of this approach and differences
to \citet{Tricco2016} are discussed in the Appendix of \citet{VarmaMuller2021}.
The extended system of MHD equations for the density $\rho$, velocity $\mathbf{v}$, magnetic field $\mathbf{B}$, the total energy density $\hat{e}$, mass
fractions $X_i$, and the rescaled Lagrange multiplier $\hat{\psi}$ reads,
\begin{eqnarray}
\partial_t \rho
+\nabla \cdot \rho \mathbf{v}
&=&
0,
\\
\partial_t (\rho \mathbf v)
+\nabla \cdot \left(\rho \mathbf{v}\mathbf{v}-
\frac{\mathbf{B} \mathbf{B}}{4\pi}
+P_\mathrm{t}\mathcal{I}
\right)
&=&
\rho \mathbf{g}
-
\frac{(\nabla \cdot\mathbf{B}) \mathbf{B}}{4\pi}
,
\\
\partial_t {\hat{e}}+
\nabla \cdot 
\left[(e+P_\mathrm{t})\mathbf{v}
-\frac{\mathbf{B} (\mathbf{v}\cdot\mathbf{B})
{-c_\mathrm{h} \hat{\psi} \mathbf{B}}}{4\pi}
\right]
&=&
\rho \mathbf{g}\cdot \mathbf{v}
+
\rho \dot{\epsilon}_\mathrm{nuc}
,
\\
\partial_t \mathbf{B} +\nabla \cdot (\mathbf{v}\mathbf{B}-\mathbf{B}\mathbf{v})
+\nabla  (c_\mathrm{h} \hat{\psi})
&=&0,
\\
\partial_t \hat{\psi}
+c_\mathrm{h} \nabla \cdot \mathbf{B}
&=&-\hat{\psi}/\tau,
\\
\partial_t (\rho X_i)
+\nabla \cdot (\rho  X_i \mathbf{v})
&=&
\rho \dot{X}_i
,
\end{eqnarray}
where $\mathbf{g}$ is the gravitational acceleration, $P_\mathrm{t}$ is the total pressure, $\mathcal{I}$ is the Kronecker tensor, $\tau$ is the damping timescale for divergence cleaning, and $\dot{\epsilon}_\mathrm{nuc}$ and $\dot{X}_i$ are energy and mass fraction
source terms from nuclear reactions. This
system conserves the volume integral of a
modified total energy density $\hat{e}$,
which also contains the cleaning field $\hat{\psi}$,
\begin{equation}
{\hat{e}}
=\rho \left(\epsilon+\frac{v^2}{2}\right)+\frac{B^2+\hat{\psi}^2}{8\pi},
\end{equation}
where $\epsilon$ is the mass-specific internal energy. 

To reduce numerical dissipation
near the grid axis in our simulations,
we have modified the mesh coarsening
algorithm of \citet{Muller2019a} by implementing
a third-order accurate slope-limited prolongation
scheme.
Neutrinos are treated using the \textsc{FMT} (fast multi-group transport)
scheme of \citet{Mueller2015}, which solves the energy-dependent
zeroth moment equation for three neutrino species
in the stationary approximation using a closure
obtained from a two-stream solution of the Boltzmann equation

The two models are run with a grid resolution of $550\times128\times 256$
zones in $r$, $\theta$ and $\varphi$ (corresponding
to $1.4^\circ$ in angle), an exponential
grid in energy spaces with $21$ zones from $4 \, \mathrm{MeV}$
to $240 \, \mathrm{MeV}$. The simulations use the equation of state of \citet{Lattimer1991} with
an incompressibility modulus of $K=220\, \mathrm{MeV}$ at high densities, and an ideal gas consisting of photons, electrons, positions, and
non-relativistic nucleons and nuclei, in conjunction with an NSE solver above $5 \, \mathrm{GK}$  and
a flashing treatment at lower temperatures \citep{Rampp2000}.

\section{Results}
\label{sec:results}

\subsection{Explosion Dynamics}
\label{subsec:dynamics}

\begin{figure}
   \centering
  \includegraphics[width=\linewidth]{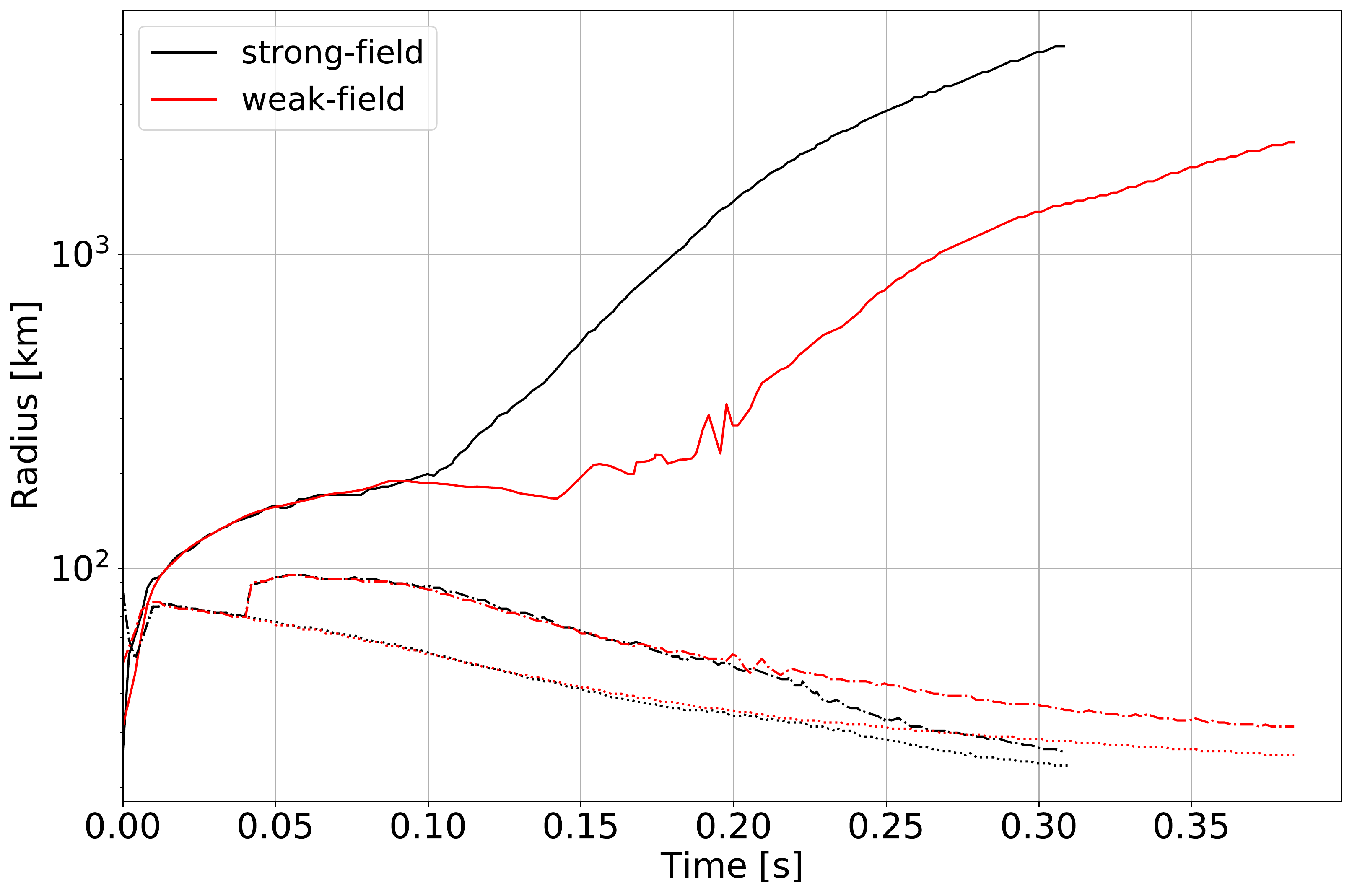}
  \caption{Evolution of the maximum
  shock radius (solid), PNS radius (dotted), and gain radius (dash-dotted) for the strong-field (black) and weak-field (red) models. 
  }
  \label{fig:Radii}
\end{figure}

\begin{figure}
   \centering
  \includegraphics[width=\linewidth]{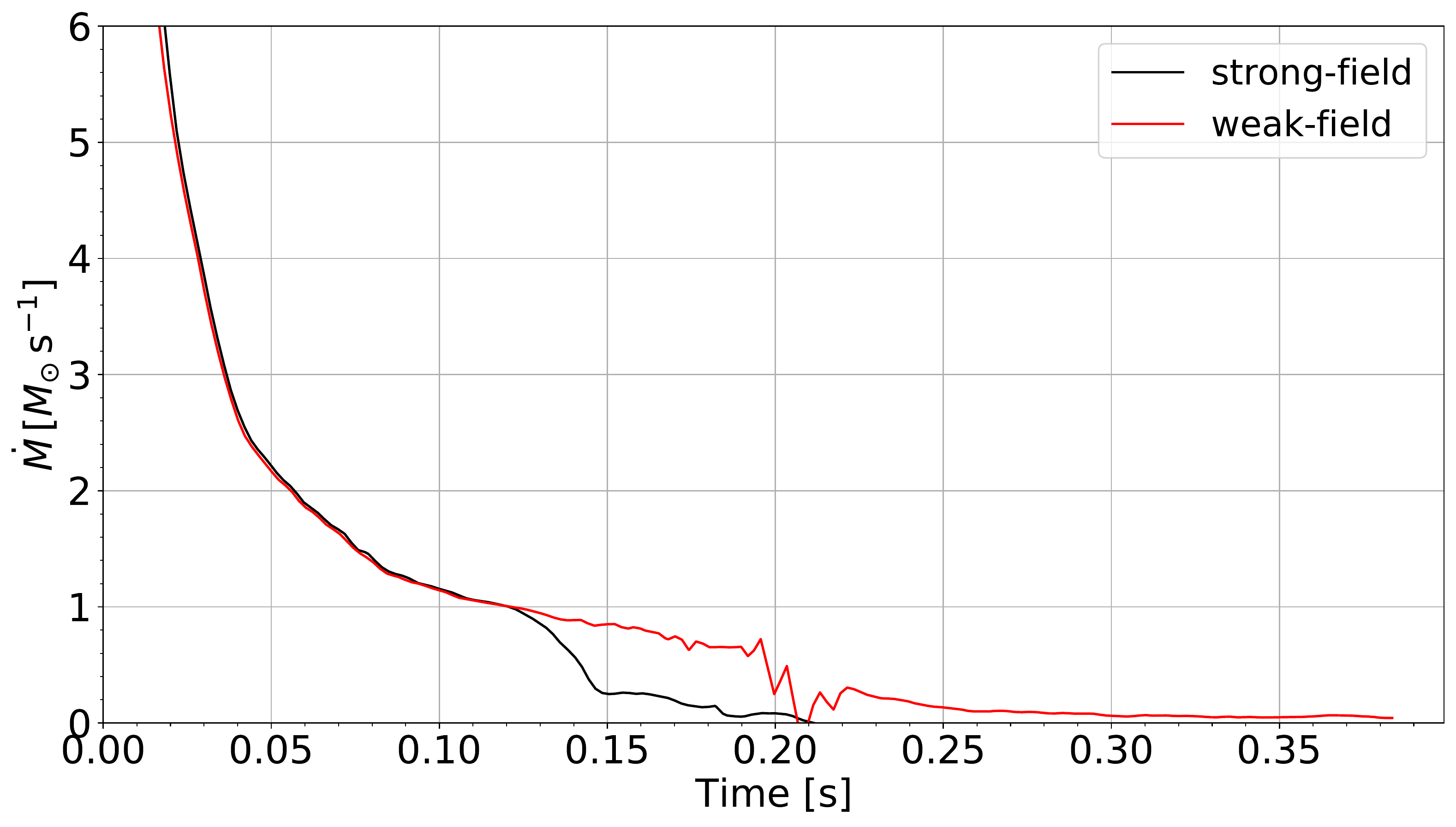}
  \caption{Mass accretion rate $\dot{M}$ for the strong-field model (black) and for the weak-field model (red), calculated at a radius of $250\,\mathrm{km}$.
  }
  \label{fig:Macc}
\end{figure}

Both the strong- and weak-field models  collapse to a PNS within the same time, $\mathord{\approx} 0.1\,\mathrm{s}$ after the models are mapped to 3D. In Figure~\ref{fig:Radii}, we see that while their initial shock trajectories are very similar, they begin to deviate significantly after about $0.1\,\mathrm{s}$ post-bounce, at which point the strong magnetic field model undergoes much more rapid shock expansion. In the weak-field case, the shock is also revived, but shock expansion sets in about $0.05\texttt{-}0.1\, \mathrm{s}$ later.
Up to the time of explosion in the strong-field model,
the mass accretion rate agrees perfectly in both runs
(Figure~\ref{fig:Macc}), indicating that the strong initial
field does not spuriously affect the
\emph{spherically-averaged} collapse dynamics. After the onset of the explosion, the accretion rate drops visibly in both cases. In the strong-field model, net accretion on the PNS \Red{(defined as the radius where the angle-averaged density reaches $10^{11}\,\mathrm{g}\,\mathrm{cm}^{-3}$)} ceases shortly after $0.2\, \mathrm{s}$ post-bounce. The subsiding accretion is responsible for a change in PNS surface structure that is reflected in the evolution of the PNS radius and the gain radius in Figure~\ref{fig:Radii}. While the two models match very closely in terms of PNS radius and gain radius for the first $0.225\, \mathrm{s}$, they then start to diverge, with the PNS radius and gain radius of the strong magnetic field models dropping more rapidly.

\begin{figure}
   \centering
  \includegraphics[width=\linewidth]{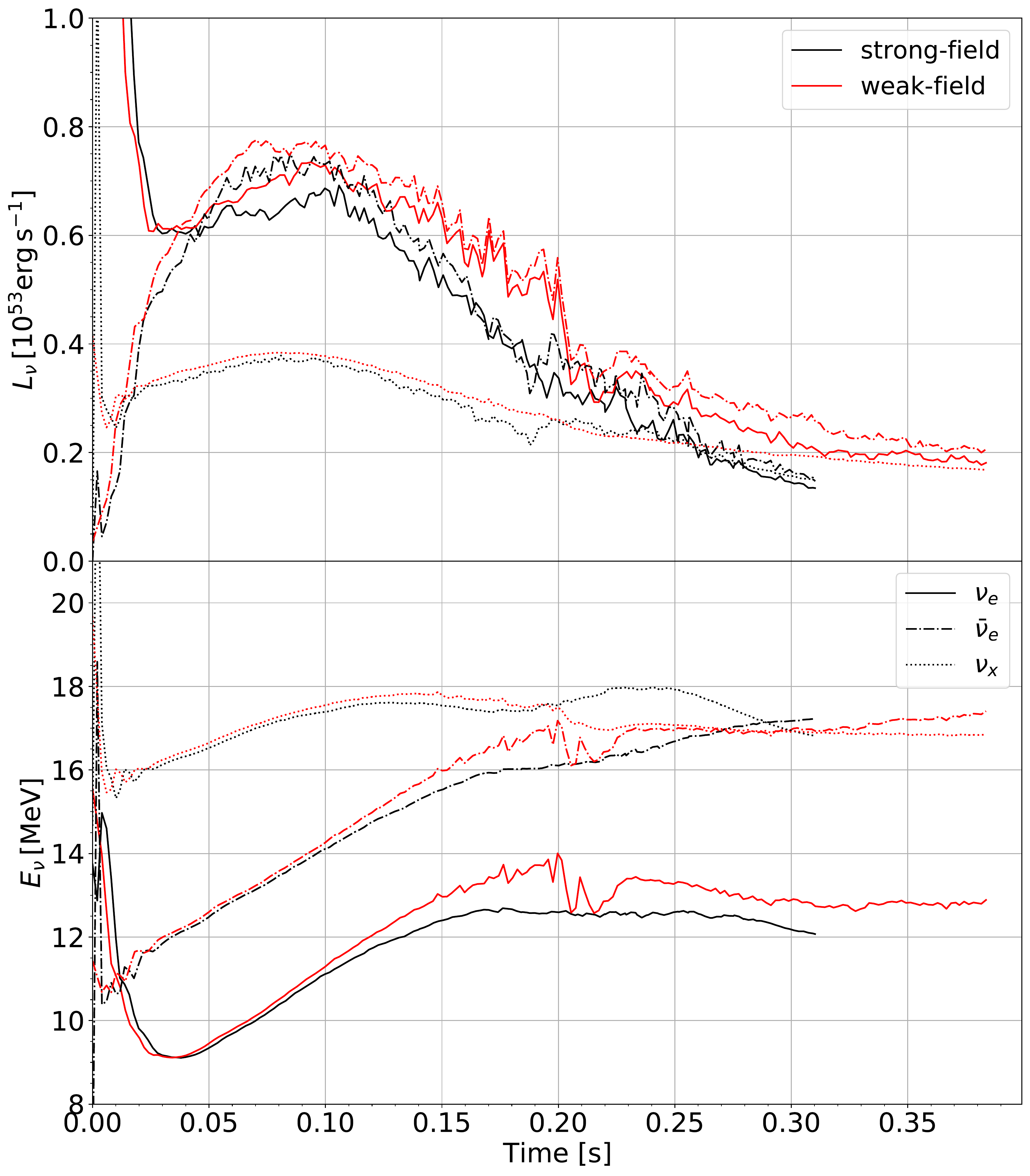}
  \caption{Neutrino luminosities (top) and mean energies (bottom) of electron neutrinos ($\nu_e$), electron antineutrinos ($\bar{\nu}_e$) and heavy-flavour neutrinos ($\nu_\mathrm{x}$), measured at a radius of $2000\,\mathrm{km}$.
  }
  \label{fig:Neutrino}
\end{figure}

Shock revival also has consequences for the neutrino luminosities and mean energies, which are shown in Figure~\ref{fig:Neutrino}. Up until $0.1\,\mathrm{s}$ after bounce, the neutrino-emission is very similar in both models. \Red{Small differences arise
because of stochastic variation in the
early post-bounce entropy profiles that are imprinted on the PNS by prompt convection 
and slightly change the thermodynamics conditions at the neutrinosphere
\citep[cp.][]{MullerVarma2020}.}
Once the shock starts to expand in the
strong-field model, the electron neutrino and
antineutrino luminosities drop significantly faster than in the weak-field model, and the rise of their
mean energies is slower. Interestingly, the heavy-flavour neutrino mean energy remains higher in the strong-field model, which is consistent with faster contraction of the PNS. It is also noteworthy that there are small differences in the neutrino emission  between the strong- and weak-field models prior to shock revival, which would actually suggest slightly better heating conditions in the weak-field model. These can be ascribed to minute differences in the collapse phase and the PNS structure at early post-bounce times that ultimately stem from the slight perturbation of the spherical collapse dynamics by the introduction of the initial fields.

\begin{figure}
   \centering
  \includegraphics[width=\linewidth]{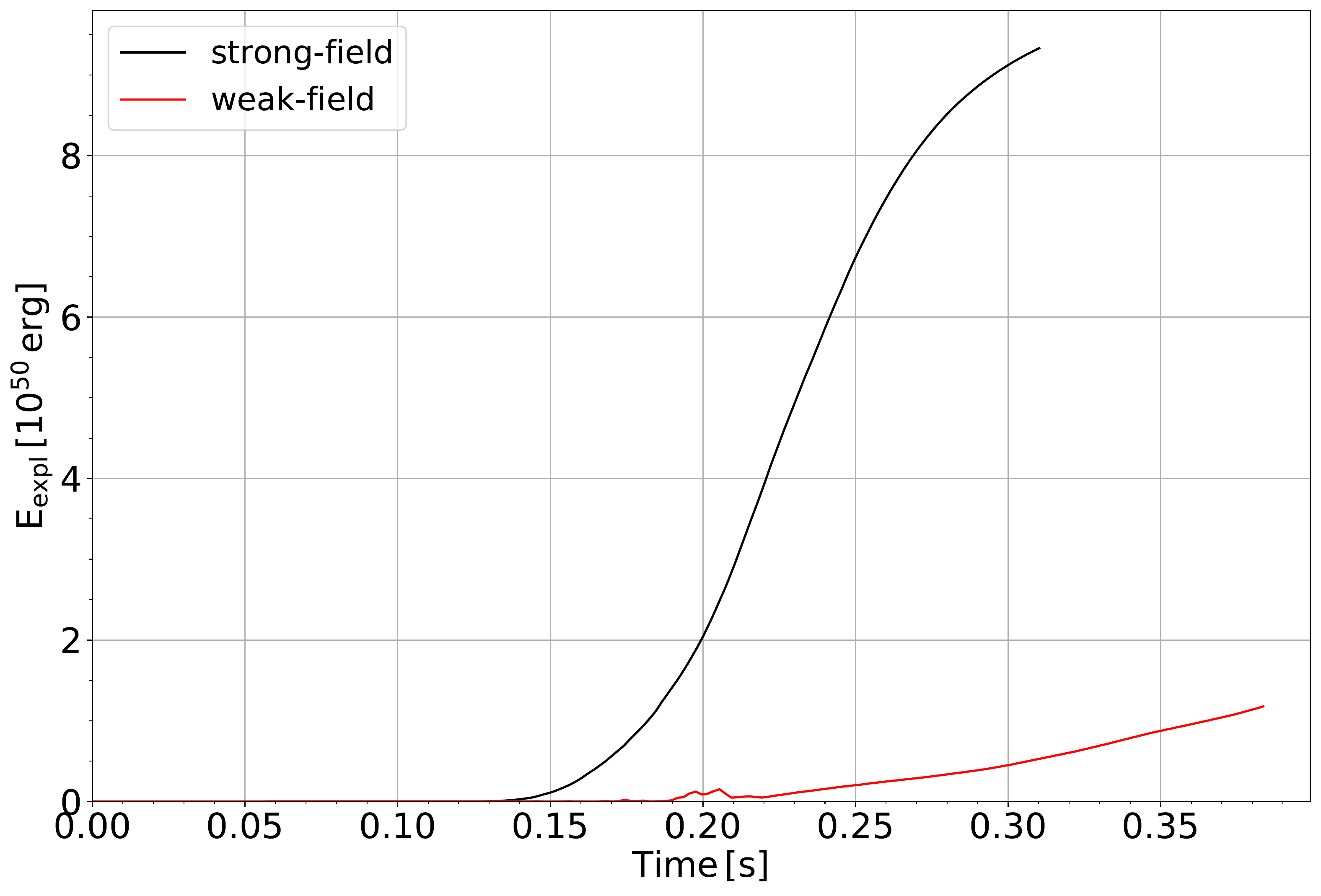}
  \caption{Evolution of the diagnostic explosion energy $E_\mathrm{expl}$ for the strong-field model (black) and  the weak-field model (red).
  }
  \label{fig:Eexp}
\end{figure}

Figure~\ref{fig:Eexp} shows the diagnostic explosion energy
$E_\mathrm{expl}$ \citep{Buras2006} for both models. The diagnostic explosion energy  is defined as an integral over the region that is nominally 
unbound,
\begin{equation}
E_\mathrm{expl} = \int\limits_{e_\mathrm{tot}>0} \rho e_\mathrm{tot}
\,\ud V ,
\end{equation}
where $e_\mathrm{tot}$ is the total energy density, i.e., the sum of the internal, kinetic, gravitational, and magnetic energy density.

\begin{figure}
   \centering
  \includegraphics[width=\linewidth]{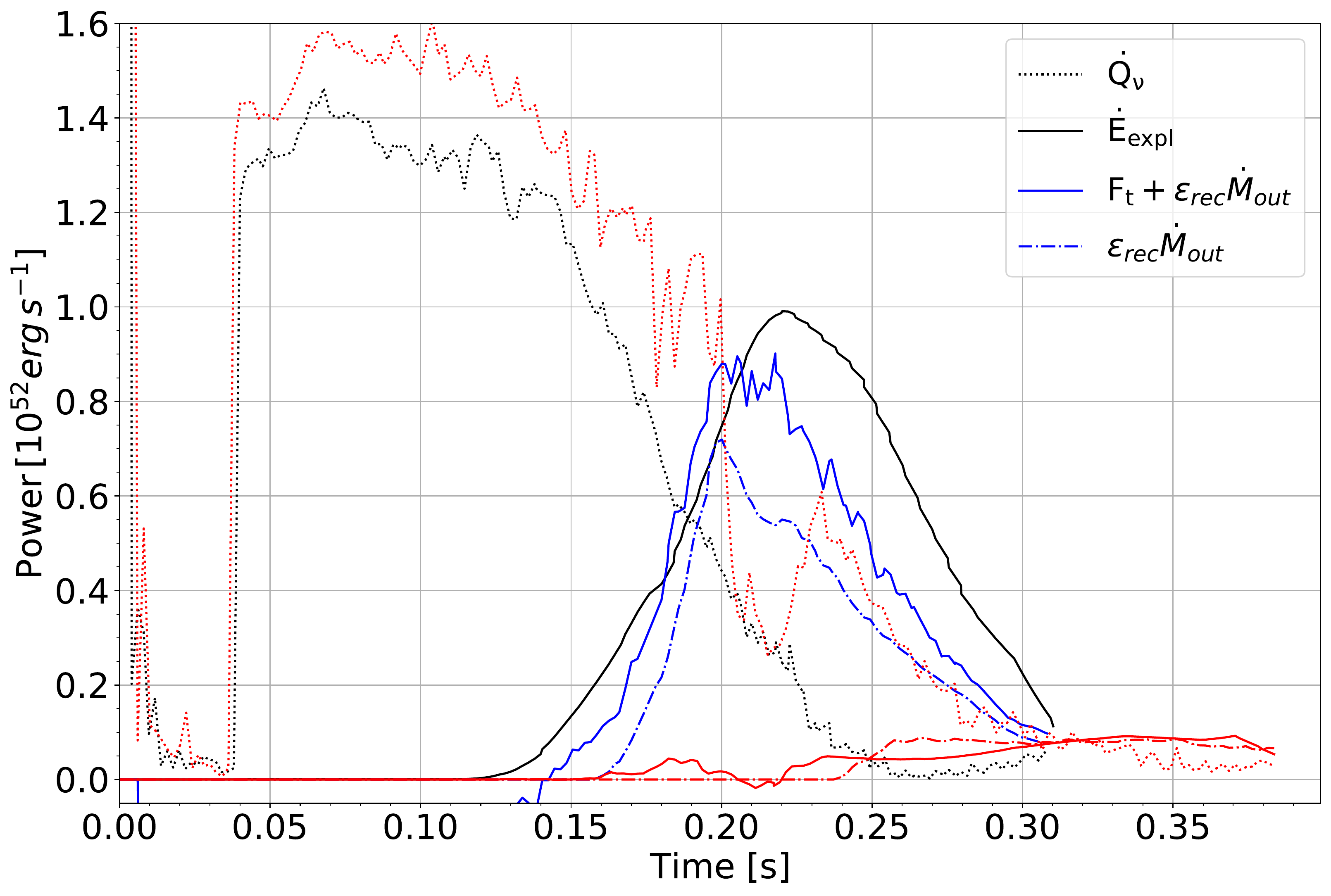}
  \caption{
  Comparison of the growth rate  ${\dot{E}_\mathrm{expl}}$
  of the diagnostic explosion energy  (solid black and red curves) to the approximation for the rate of
  energy ejection into the ejecta from
  Equation~ (\ref{eq:dE_Mdot}) (solid blue curve), which 
  accounts for the release of
  $\epsilon_\mathrm{rec}= 6\, \mathrm{MeV}/\mathrm{baryon}$
  by nucleon recombination for the measured mass outflow rate ${\dot{M}_\mathrm{out}}$ and the 
  measured magnetohydrodynamic energy flux $F_\mathrm{t}$ at the base of the gain region of the strongly magnetised model (blue). The nuclear recombination power $\epsilon_\mathrm{rec} \dot{M}_\mathrm{out}$ is also plotted on its own for both models (dash-dotted blue and red curves).  The volume-integrated neutrino heating rate $\dot{Q}_{\nu}$ is also shown (dotted curves).
  }
  \label{fig:Eexp_growth}
\end{figure}

The diagnostic explosion energy of the strong-field model begins to rise around the time when the shock trajectories of the two models deviate, and reaches more tan $9\times10^{50}\,\mathrm{erg}$ by the end of the simulation. The explosion energy already appears to level off at this stage, which is rather unusual when compared to current 3D hydrodynamic simulations of core-collapse supernova explosions. Long-term 3D explosion models of massive progenitors typically show a sustained rise of the explosion energy over up to several seconds \citep{Muller2015a,Bruenn2016,bollig_21}, but at a significantly smaller growth rate, more reminiscent of the weak-field model, which has only reached about
$10^{50}\, \mathrm{erg}$. The unusually rapid growth and saturation
of the explosion energy is further emphasised by Figure~\Red{\ref{fig:Eexp_growth}}, which shows that the rate of growth
$\ud E_\mathrm{expl}/\ud t$ peaks at $10^{52}\, \mathrm{erg}\, \mathrm{s}^{-1}$ about $0.22\, \mathrm{s}$ post-bounce, and drops by more than a factor of five afterwards.

\begin{figure}
   \centering
  \includegraphics[width=\linewidth]{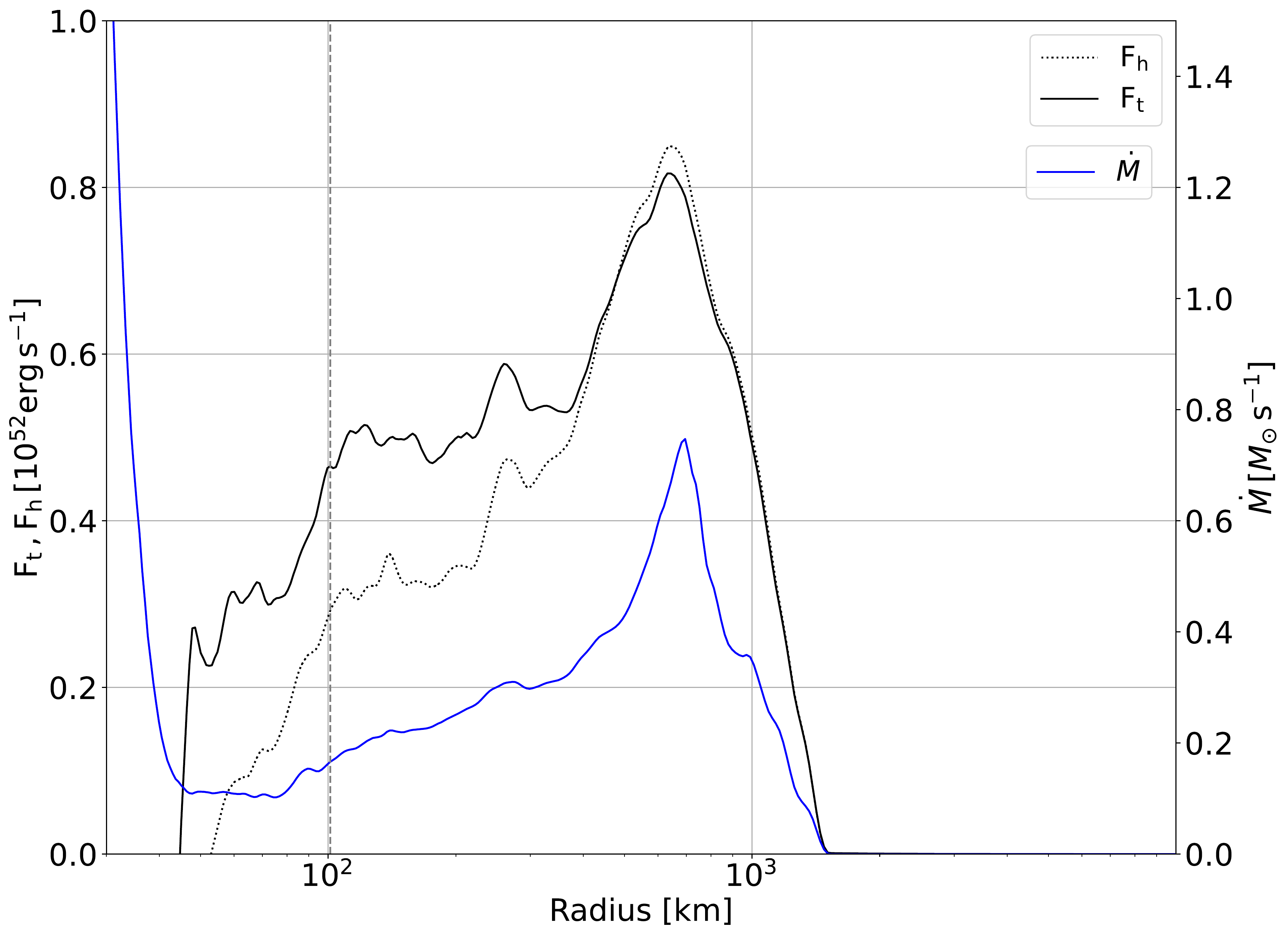}
  \caption{Radial profiles 
  of the total magnetohydrodynamic energy flux $F_\mathrm{t}$ with rest masses subtracted (black, solid),
  the purely hydrodynamic total enthalpy flux
  $F_\mathrm{h}$ (black, dotted), and the mass outflow rate
  $\dot{M}_\mathrm{out}$ (blue) for the strong-field model at $0.20\,\mathrm{s}$ post-bounce. The energy and mass fluxes are computed as integrals over the area covered by outflows with $v_r>0$. The vertical dashed line at $\approx \Red{100}\,\mathrm{km}$ indicates where does the angle-averaged total energy in the outflows become positive.
  }
  \label{fig:EFlux}
\end{figure}

Strong initial fields thus evidently have a major impact on the explosion dynamics, allowing the model
to reach a typical supernova explosion energy
\citep{kasen_09,murphy_19} within only $0.3\, \mathrm{s}$
after bounce. As we shall show, this effect is largely of an indirect nature. It comes about because the strong initial fields help trigger the explosion early, rather than from direct delivery of the explosion energy by magnetic stresses as in the magnetorotational paradigm. The explosion in energised in a similar manner as in existing hydrodynamics models of neutrino-driven explosions
with a few subtle differences related to early shock revival.

Previous studies by \citet{Marek2006,Muller2015a, Bruenn2016} have shown that for the neutrino-driven explosion mechanism, the growth of the diagnostic explosion energy is largely determined by the mass outflow rate, ${\dot{M}_\mathrm{out}}$ of the ejecta, which are first lifted to $e_\mathrm{tot} \approx 0$ by neutrino heating, before obtaining most of their net positive energy from nucleon recombination. Ignoring secondary effects like the scooping-up of bound material by the shock and losses of explosion energy by turbulent braking of the expanding neutrino-heated bubbles, this results in a growth rate of the explosion energy of
$\dot{E}_\mathrm{expl}\approx \epsilon_\mathrm{rec} \dot{M}_\mathrm{out}$, where $\epsilon_\mathrm{rec}$
is the recombination energy per unit mass.

Comparing the rate of energy release $\epsilon_\mathrm{rec} \dot{M}_\mathrm{out}$ by nucleon recombination to the rate of growth of
the explosion energy is rather tricky for the strong-field model since $\dot{M}_\mathrm{out}$ is non-stationary and varies
strongly with radius (Figure~\ref{fig:EFlux}), unlike
in models with later shock revival where 
$\dot{M}_\mathrm{out}\approx \mathrm{const.}$ over a large range in radius (see, e.g., Figure~18 of \citealt{Muller2015a}). To capture the energy release by nuclear recombination with reasonable accuracy during a rather non-stationary outflow, it is critical to evaluate
$\dot{M}_\mathrm{out}$ not too far away from the actual recombination radius.  In our subsequent analysis, we evaluate $\dot{M}_\mathrm{out}$ at a radius of $400\,\mathrm{km}$ by integrating over all directions where the radial velocity component $v_r$ is positive,
\begin{equation}
\dot{M}_\mathrm{out} = \int\limits_{v_r>0} \rho v_r r^2
\,\ud \Omega.
\end{equation}
Even in this case, however, nucleon recombination can only account for a little more than half of the growth rate of the explosion energy. 
% \fabian{It is not obvious that the dot-dashed curve in Fig 6 integrates to about half of that of the solid line. Can we add the integral of dotM*eps to Fig. 5?}

The extra energy input into the explosion can be
explained by $P\,\mathrm{d}V$ work
(and to some extent, work by magnetic stresses) by the underlying layers on the ejecta region.
It is noteworthy that a more significant role of
the magnetohydrodynamic energy flux into the ejecta region
compared to recombination was also found for early explosions in axisymmetry \citep{Bruenn2016} and for
some mass shells in rapidly developing electron-capture supernova explosions \citep{janka_08}.

Taking into account the hydrodynamic energy flux into the ejecta region we expect the explosion energy to grow as
\begin{equation}
\frac{\ud E_\mathrm{expl}}{\ud t}  \approx
F_\mathrm{t} + \epsilon_\mathrm{rec} \dot{M}_\mathrm{out}, 
\label{eq:dE_Mdot}
\end{equation}
where $F_\mathrm{t}$ is the total magnetohydrodynamic energy flux,
\begin{equation}
F_\mathrm{t} = \int\limits_{v_r>0}  r^2 
\left[(\rho e+P_\mathrm{t}+\rho \Phi) v_r + \frac{\mathbf{B} (\mathbf{v}\cdot\mathbf{B})}{4\pi} \right]
\,\ud \Omega .
\end{equation}
In Figure~\ref{fig:Eexp_growth}, we compare 
$F_\mathrm{t}+\epsilon_\mathrm{rec} \dot{M}_\mathrm{out}$ to $\dot{E}_\mathrm{expl}$ using a value of  $\epsilon_\mathrm{rec} = 6\, \mathrm{MeV}/m_\mathrm{b}$ (where $m_\mathrm{b}$ is the atomic mass unit), which accounts for incomplete recombination as well as some turbulent mixing. 
The growth of ${E}_\mathrm{expl}$ is described reasonably well by the assumption of delivery by nuclear recombination and an additional, subdominant contribution by $P\,\ud V$-work and magnetic stresses, although  $\epsilon_\mathrm{rec} \dot{M}_\mathrm{out}$ alone somewhat underestimates the actual growth rate. 

To quantify the direct role of magnetic fields in
powering the explosion, we also compare the total magnetohydrodynamic energy flux $F_\mathrm{t}$ to the purely hydrodynamic total enthalpy flux $F_\mathrm{h}$,
\begin{equation}
F_\mathrm{h} = \int\limits_{v_r>0} r^2 
(\rho e+P+\rho \Phi) v_r 
\,\ud \Omega,
\end{equation}
where the magnetic pressure and magnetic stresses are excluded and only
the gas pressure $P$ enters aside from purely advective terms.
At the base of the ejecta region at
\Red{$100\,\mathrm{km}$} (where the angle-averaged total energy in the outflows becomes positive), the
% hydrodynamic enthalpy flux clearly dominates over
% the residual work done by magnetic pressure and magnetic stresses.
\Red{hydrodynamic enthalpy flux $F_\mathrm{h}$ accounts for most of the total energy flux $F_\mathrm{t}$ into the ejecta. The rest of the energy flux into the ejecta is due to work by magnetic pressure and magnetic stresses. Their contribution is non-negligible, but smaller than
the hydrodynamics flux $F_\mathrm{h}$ at the base of the ejecta region.}

The rate of growth of the explosion energy is therefore well explained by nuclear recombination and some additional magnetohydrodynamic energy flux into the ejecta region. The unusually rapid growth of the explosion energy is merely the result of a rather high mass outflow rate, and this in turn is simply due to the early onset of the explosion at a time when the mass accretion rate is still high (close to $1 M_\odot\, \mathrm{s}^{-1}$) and the mass of the gain region is large. At first glance, the fact that 
$\dot{E}_\mathrm{expl}$ even exceeds the volume-integrated neutrino heating rate $\dot{Q}_\nu$ in the gain region (Figure~\ref{fig:Eexp_growth}) might suggest that additional powering other than neutrino heating is needed to explain the high mass outflow rate. For a steady-state outflow one would expect
a ratio $\dot{E}_\mathrm{expl}/\dot{Q}_\nu\approx \epsilon_\mathrm{nuc}/|e_\mathrm{bind}|$,
where $|e_\mathrm{bind}|$ is the typical binding energy of neutrino-heated matter before it turns away from the PNS to be ejected, and if $|e_\mathrm{bind}|$ is taken as the net binding energy at the gain radius, one usually obtains $\dot{E}_\mathrm{expl}/\dot{Q}_\nu<1$.
The rather small magnetohydrodynamic energy flux
of
only $\mathord{\sim} 3\times 10^{51}\, \mathrm{erg}\, \mathrm{s}^{-1}$ deeper in the gain region at radii of less than $100\, \mathrm{km}$ in Figure~\ref{fig:EFlux} argues against an additional power source for lifting the ejecta to roughly zero net energy before recombination. The large ratio $\dot{E}_\mathrm{expl}/\dot{Q}_\nu$ is instead explained by other effects. First, \citet{Mueller_2017} pointed out that the effective turnaround radius of the ejecta can become quite large in a developing explosion, so that material need not be lifted out of the potential well all the way from the gain radius. Second, the average binding energy in the gain \Red{region} decreases quite quickly in the strong-field model
as shown in Figure~\ref{fig:Ebind_gain}. Third, for a non-stationary outflow, $\dot{E}_\mathrm{expl}$
and $\dot{Q}_\nu$ cannot be compared at the same time
because there is a delay between peak neutrino heating of an ejected mass element in the gain region and the time when it acquires contributes to the explosion energy after undergoing recombination. 

The dynamics of the outflows in the strong-field model thus clearly point towards an \emph{indirect} effect of the initial field on the explosion energy. The key role of the magnetic fields is to \emph{trigger} shock revival early, which provides the conditions for a rapid growth of the explosion energy. We therefore need to elucidate how the strong initial fields precipitate shock revival.

\subsection{Dynamics of the Gain Region and Evolution Towards Shock Revival}
\label{subsec:gaindynamics}
Contrary to the explosion energetics, the strong initial magnetic fields play a clearly recognisable role in the dynamics of the gain region prior to and around shock revival. The earlier onset of explosion in the strong-field model appears to be partly due to similar effects as in \citet{MullerVarma2020}, where the magnetic fields in the gain region were generated by a turbulent dynamo, but our analysis also reveals a new phenomenon that is peculiar to stellar cores with  strong pre-collapse fields\Red{.}

\begin{figure}
   \centering
  \includegraphics[width=\linewidth]{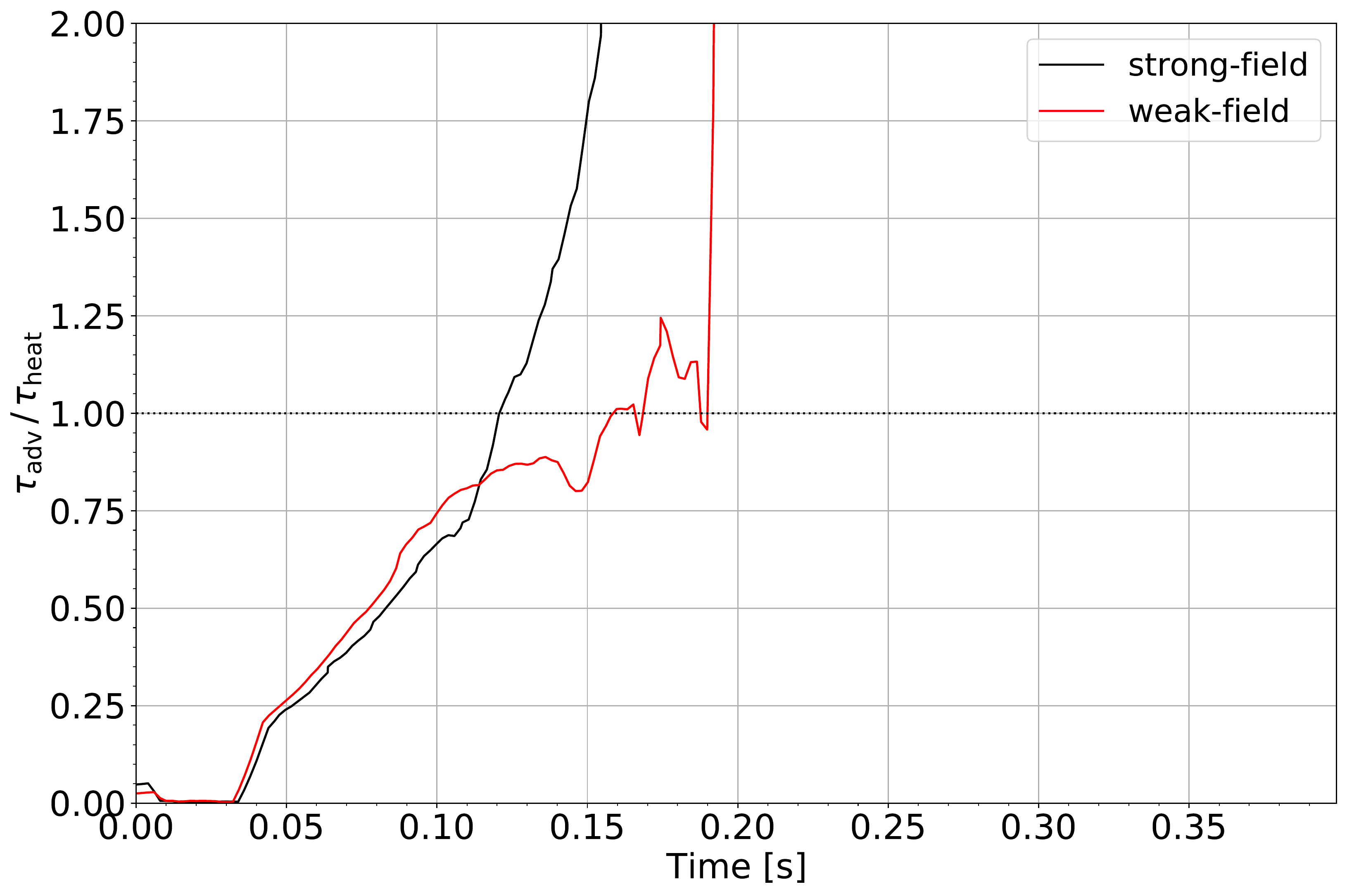}
  \caption{Ratio of the advection timescale
  $\tau_\mathrm{adv}$ to the heating timescale
  $\tau_\mathrm{heat}$  in the gain region of the strong-field model (black) and the weak-field model (red). A horizontal dotted line indicates the threshold criterion $\tau_\mathrm{adv}/\tau_\mathrm{heat}=1$
  for runaway shock expansion in the neutrino-driven mechanism.
  }
  \label{fig:Tau}
\end{figure}

To analyse the dynamics of the gain region, we first consider the critical ratio of the advection and heating time scales $\tau_\mathrm{adv}/\tau_\mathrm{heat}$, (Figure~\ref{fig:Tau}),  which quantifies the conditions for runaway shock expansion by neutrinos \citep{Buras2006}. The advection and heating timescale  $\tau_\mathrm{adv}$  and $\tau_\mathrm{heat}$  are
defined as
\begin{equation}
\tau_\mathrm{adv} = \frac{M_\mathrm{g}}{\dot{M}},
\end{equation}
\begin{equation}
\tau_\mathrm{heat} = \frac{|E_\mathrm{g}|}{\dot{Q}_\nu},
\end{equation}
where $E_\mathrm{g}$ and $M_\mathrm{g}$ are the binding energy and mass of the gain region, $\dot{M}$ is the mass accretion rate, and $\dot{Q_\nu}$ is the volume-integrated neutrino heating rate.

It is noteworthy that by the time the shock starts
to expand in the strong-field model and deviates
from the shock trajectory in the weak-field model
at $0.1\, \mathrm{s}$ after bounce, $\tau_\mathrm{adv}/\tau_\mathrm{heat}$ still has not reached the critical threshold $\tau_\mathrm{adv}/\tau_\mathrm{heat}\approx 1$ and is actually slightly \emph{lower} than in the weak-field model. The lower ratio $\tau_\mathrm{adv}/\tau_\mathrm{heat}$ can be explained by the slightly lower neutrino heating rates in the strong-field model (Figure~\ref{fig:Eexp_growth}; cp.\ the discussion of the neutrino luminosities in Section~\ref{subsec:dynamics}).
While  $\tau_\mathrm{adv}/\tau_\mathrm{heat}\approx 1$
is reached in the strong-field model about $0.02\, \mathrm{s}$ later, this provides strong evidence that dynamical effects of magnetic fields initially aid shock expansion 
(even though their importance as a subsidiary driver of the explosion diminishes once the runaway criterion is met).

\begin{figure}
   \centering
  \includegraphics[width=\linewidth]{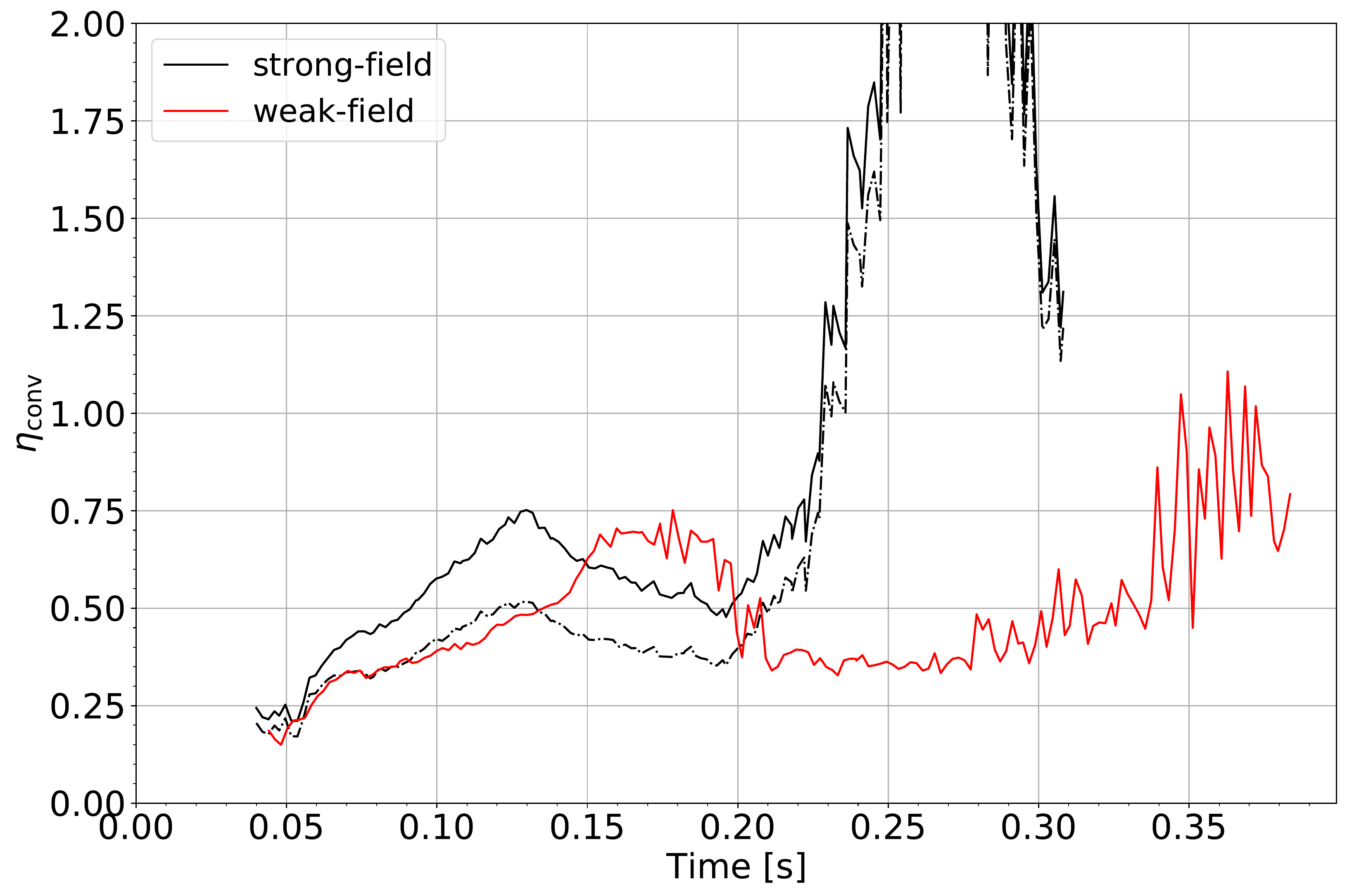}
  \caption{Evolution of the convective efficiency $\eta_\mathrm{conv}$ in the gain region as defined by Equation~(\ref{eq:Nconv}). The solid lines show the convective efficiency based on the
  total kinetic and magnetic turbulent energy
  $E_\mathrm{kin} + E_\mathrm{mag}$, while only the turbulent kinetic energy $E_\mathrm{kin}$ is included for   dash-dotted lines. Black and red curves are used for the strong- and weak-field model, respectively.
  Only the solid line is visible for the weak-field model because the magnetic energy in the gain region is negligible.
  }
  \label{fig:Nconv}
\end{figure}

We evidence for important dynamical effects of the strong seed fields both inside the gain region (as in \citealt{MullerVarma2020})  and outside the shock.
Their impact inside the gain region can be gauged by the
convective efficiency parameter for the conversion of neutrino heating to turbulent kinetic and magnetic energy in the gain region,
\citep{Muller2015b,MullerVarma2020},
\begin{equation}
    \eta_\mathrm{conv} = \frac{(E_\mathrm{kin} + E_\mathrm{\mathbf{B }})/M_\mathrm{gain}}{[(r_\mathrm{sh}-r_\mathrm{gain})\dot{Q}_{\nu}/M_\mathrm{gain}]^{2/3}},
    \label{eq:Nconv}.
\end{equation}
As shown by Figure~\ref{fig:Nconv}, $\eta_\mathrm{conv}$ is
$\mathord{\lesssim} 50\%$ higher than in the weak-field model until about $0.1255\,\mathrm{s}$ post-bounce. The convective efficiency computed from the turbulent kinetic energy alone roughly equals the total convective efficiency in the weak-field model, in which the turbulent magnetic energy is negligible during this phase. The situation is thus analogous to that in \citet{MullerVarma2020}; the turbulent magnetic energy is filled as an extra energy reservoir on top of the kinetic turbulent energy, which is fed by the neutrino heating. Different from \citet{MullerVarma2020}, the increase in convective efficiency is observed as soon as convection develops in the strong-field model, however.

The early boost to the convective efficiency is due to the strong seed fields, which allow the magnetic fields in the gain region to reach levels close to equipartition with little amplification, compared to the situation with weak seed fields, where 
this takes an appreciable amount of time despite exponential amplification by a turbulent dynamo \citep{MullerVarma2020}.
This is illustrated by Figure ~\ref{fig:Gain_energy}, which shows the turbulent kinetic energy
$E_\mathrm{kin}$ and the magnetic
field energy $E_B$ in the gain region.
$E_\mathrm{kin}$ and $E_\mathrm{kin}$ are computed as
\begin{align}
E_\mathrm{kin}&=\int \frac{1}{2}\rho |\mathbf{v}'|^2 \, \ud V,
\label{eq:gainenergy1}
\\
E_B&=\int \frac{|\mathbf{B}|^2}{8\pi}  \,\ud V,
\label{eq:gainenergy2}
\end{align}
where $\mathbf{v'}$ denotes the fluctuations of the velocity
around its spherical Favre average \Red{(i.e. $\mathbf{v'} = \mathbf{v} - \langle \mathbf{v} \rangle$)}. We also show the ratio
$E_B/E_\mathrm{kin}$ of the magnetic and kinetic turbulent energy,  in Figure~\ref{fig:Gain_energy_ratio}.
The strong initial magnetic field strength model undergoes a very brief exponential magnetic field amplification process, placing the ratio of $E_B/E_\mathrm{kin} \approx 40\%$ when runaway expansion occurs, similar to the saturation value
found in \citet{MullerVarma2020}. By contrast, 
although the weak magnetic field model undergoes a rapid magnetic field amplification process in the gain region after $\mathord{\approx} 0.20\, \mathrm{s}$, the neutrino-driven mechanism launches the shock before the magnetic energies are strong enough to affect the dynamics of the gain region. 
\Red{We note that there is rough energy equipartition between the radial magnetic fields (generated predominantly by radial stretching of field lines) and the non-radial components (generated predominantly by non-radial shear flows). The energies
$E_{B,r}$ and $E_{B,\theta\varphi}$ in the radial and non-radial components,
\begin{align}
E_{B,r}&=\int \frac{B_r^2}{8\pi}  \,\ud V,
\\
E_{B,\theta\varphi}&=\int \frac{B_\theta^2+B_\varphi^2}{8\pi}  \,\ud V,
\end{align}
are shown in Figure~\ref{fig:Gain_energy_ratio}, as a ratio with the kinetic energy in the gain region. For approximate equipartition, the dynamical effect of magnetic fields on shock propagation
through tension and pressure \emph{forces} is arguably quite neutral, but storing extra \emph{energy} in magnetic fields will still make it easier to unbind material and release additional energy back into the neutrino-heated material as it expands. The detailed  dynamical impact of the magnetic fields on the quasi-hydrostatic structure of the gain region still remains to be investigated more rigorously; we recall that the chain of
effects through which turbulence facilitates
shock expansion is highly non-trivial even in the hydrodynamics case
\citep[e.g.,][]{Mabanta2018}.
}

\begin{figure}
   \centering
  \includegraphics[width=\linewidth]{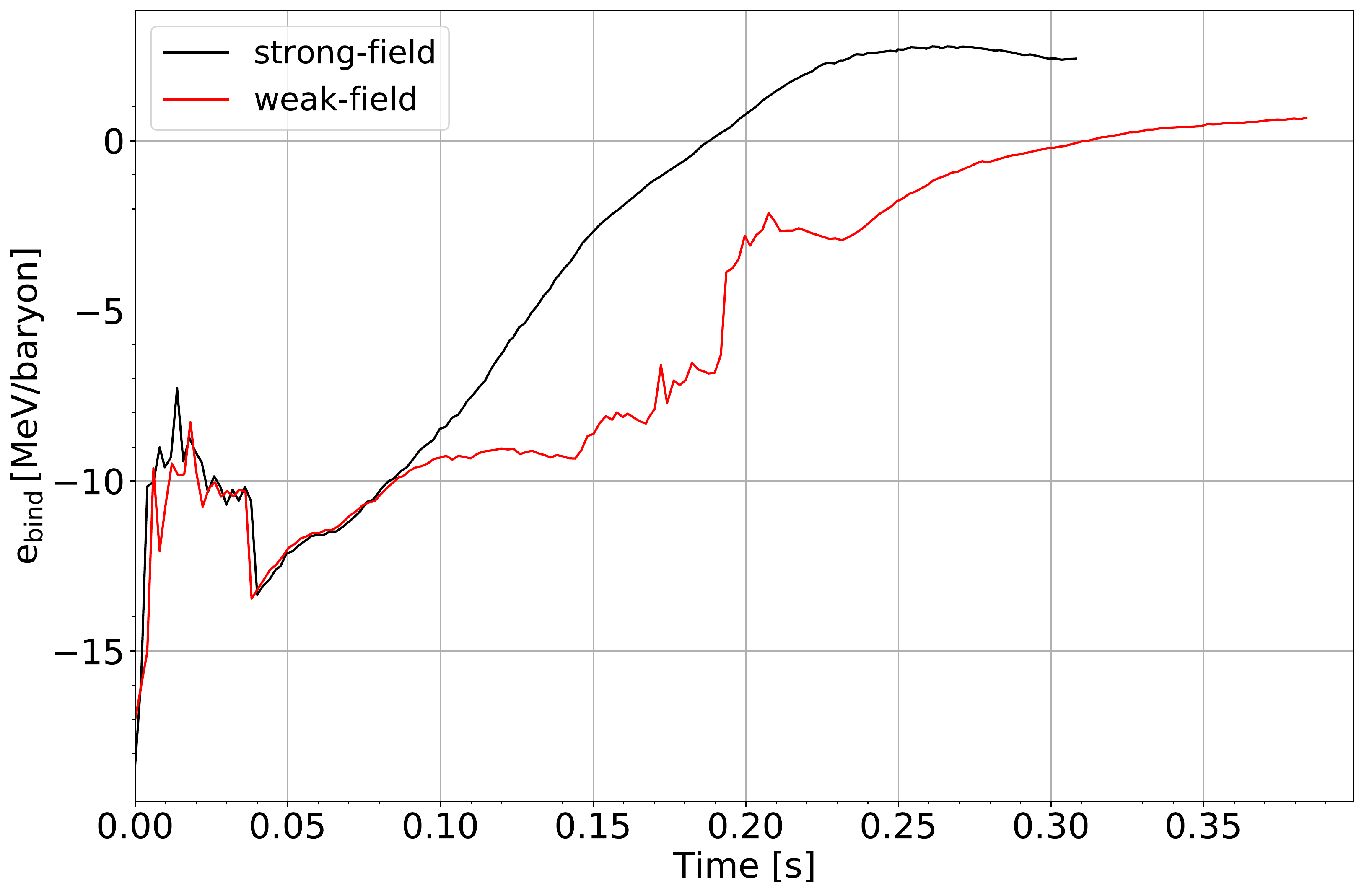}
  \caption{Average specific binding energy $e_\mathrm{bind}$ in the entire gain region of the strong-field model (black) and the weak-field (model) black.
  }
  \label{fig:Ebind_gain}
\end{figure}

\begin{figure}
   \centering
  \includegraphics[width=\linewidth]{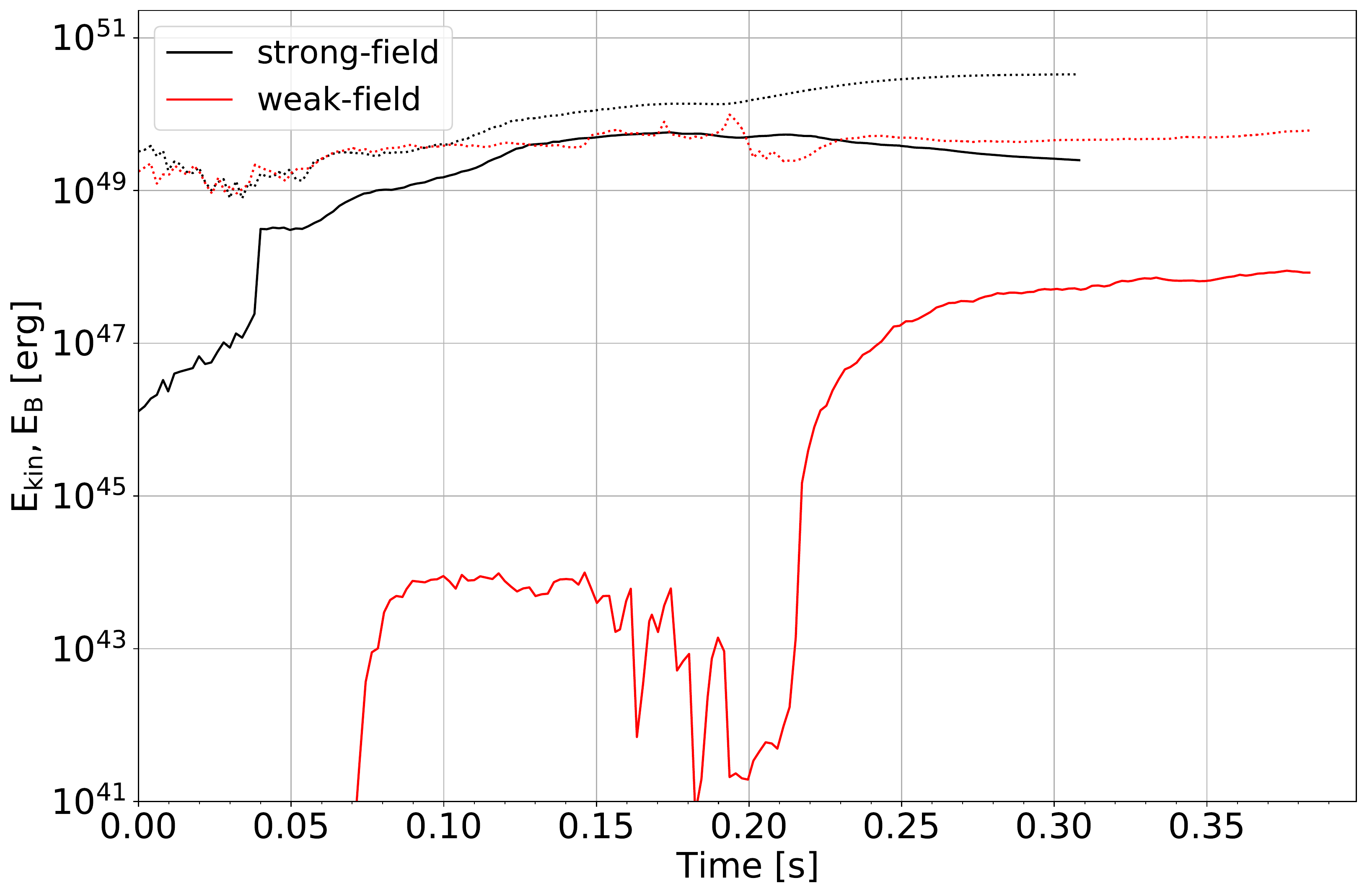}
  \caption{Evolution of the total magnetic (solid) and turbulent kinetic (dotted) energy within the gain region (Equations~\ref{eq:gainenergy1} and \ref{eq:gainenergy2}) for the strong-field model (black) and the weak-field model (red).
  }
  \label{fig:Gain_energy}
\end{figure}

\begin{figure}
   \centering
  \includegraphics[width=\linewidth]{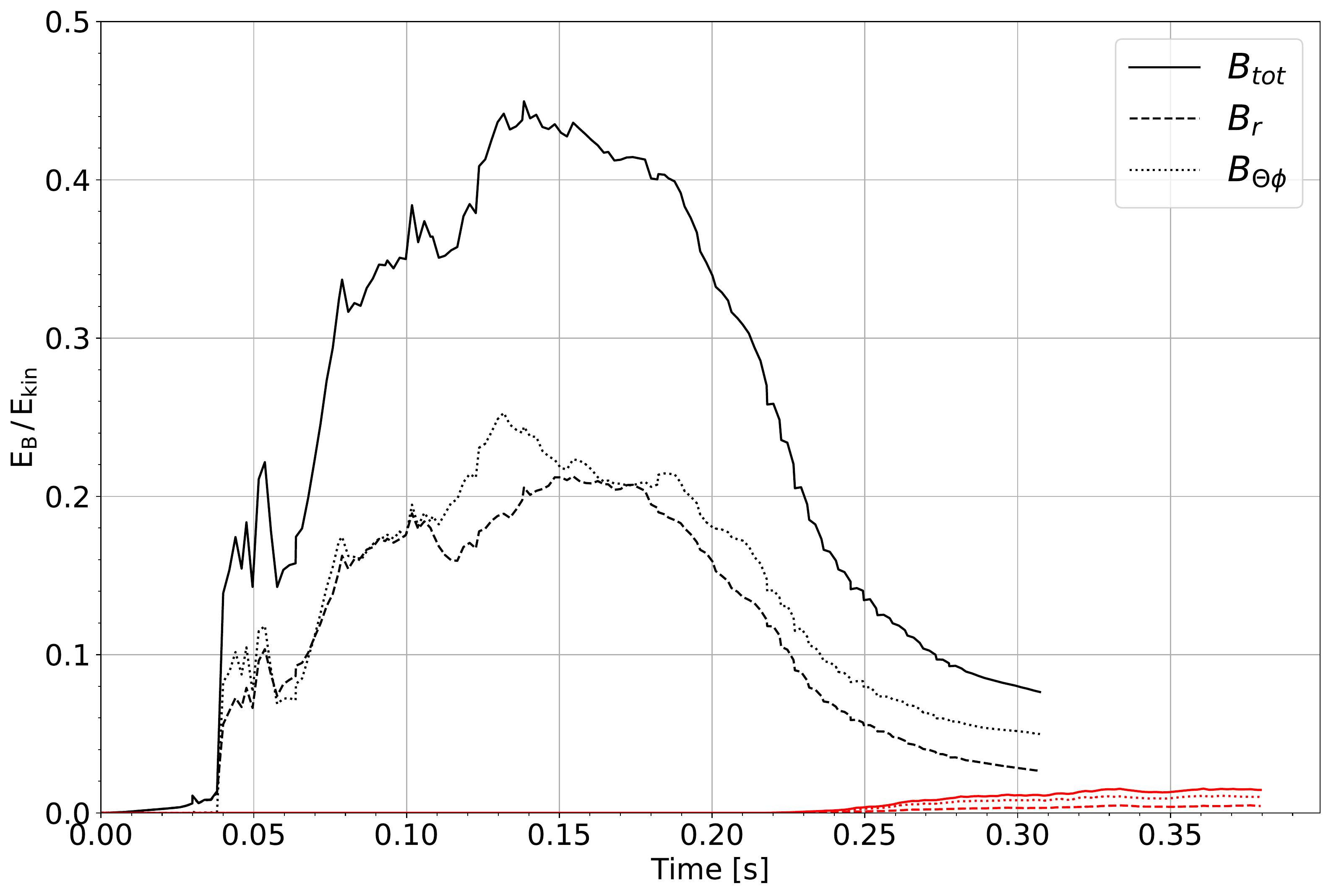}
  \caption{Evolution of the ratio between the total magnetic and turbulent kinetic  energy within the gain region for strong-field model (black) and the weak-field model (red). \Red{For both models, the ratio is plotted for the total magnetic energy (solid), the radial only component (dashed), and the angular components (dotted) of the magnetic energy.}
  }
  \label{fig:Gain_energy_ratio}
\end{figure}

In summary, a similar mechanism operates in the strong-field model
as in the MHD-aided explosion of \citet{MullerVarma2020}; magnetic fields in the gain region become sufficiently strong to boost the overall turbulent energy in the gain region and thereby aid shock revival. The key difference is that this process occurs significantly earlier due to the strong seed fields. As in \citet{MullerVarma2020}, it is not possible to disentangle precisely which turbulent terms (e.g., the magnetic stresses in the momentum equation, the flux terms in the energy equation, etc.) in the MHD equations are responsible for precipitating shock expansion. Based on the ratio $E_B/E_\mathrm{kin}$, the magnetic stress terms are clearly non-negligible. There is somewhat better evidence than in \citet{MullerVarma2020} that magnetic fields also palpably reduce the binding energy of the gain region (Figure \ref{fig:Ebind_gain}). $E_\mathrm{bind}$ in the strong-field models starts to visibly deviate from the weak-field model
from $\approx 0.09\, \mathrm{s}$ after bounce, i.e. already slightly before the shock trajectories diverge.

\begin{figure*}
\centering
    \subfloat[Strong:0.01s]{\includegraphics[width=0.45\linewidth]{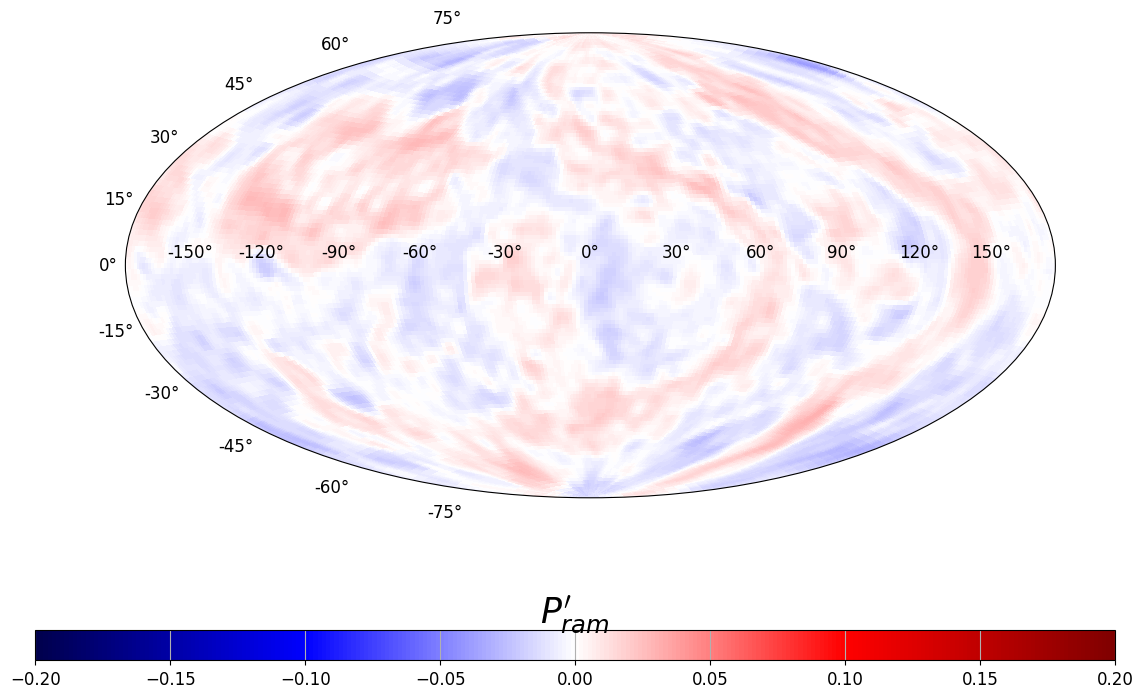}}
\hfil
    \subfloat[Weak:0.04s]{\includegraphics[width=0.45\linewidth]{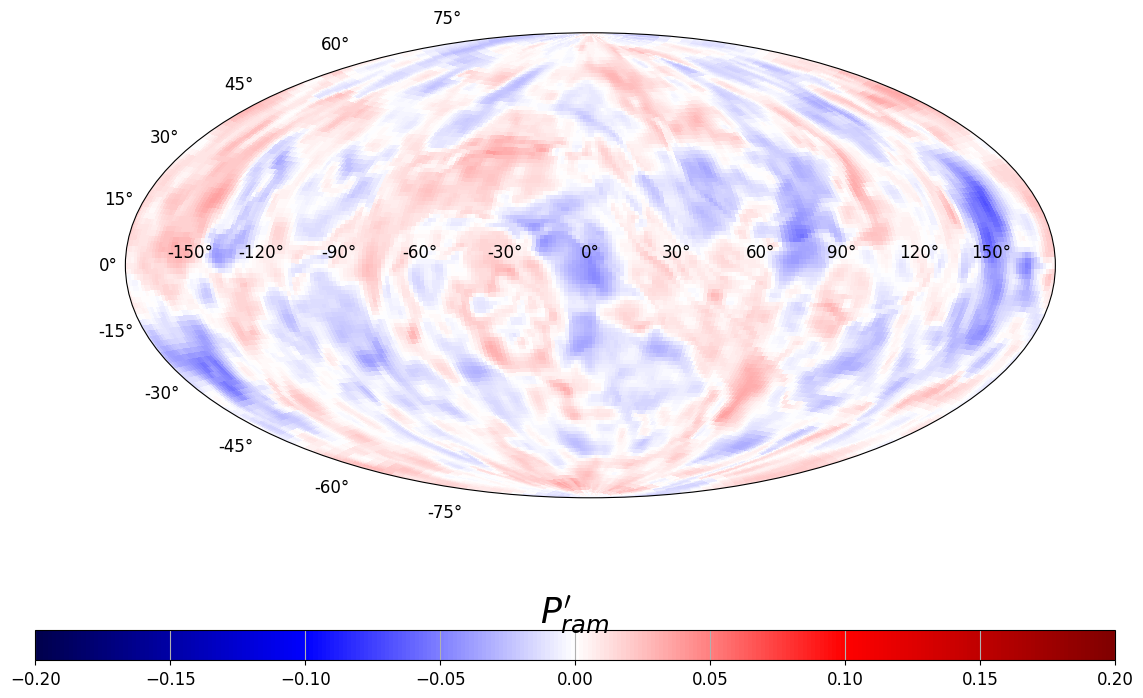}}

    \subfloat[Strong:0.05s]{\includegraphics[width=0.45\linewidth]{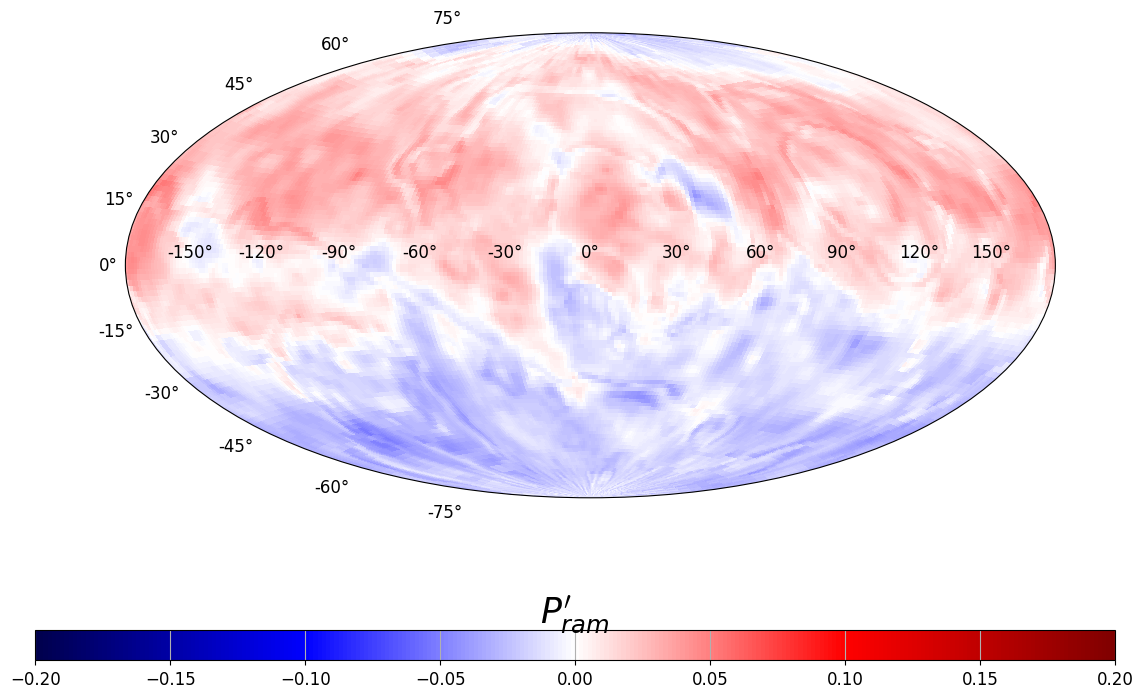}}
\hfil
    \subfloat[Weak:0.10s]{\includegraphics[width=0.45\linewidth]{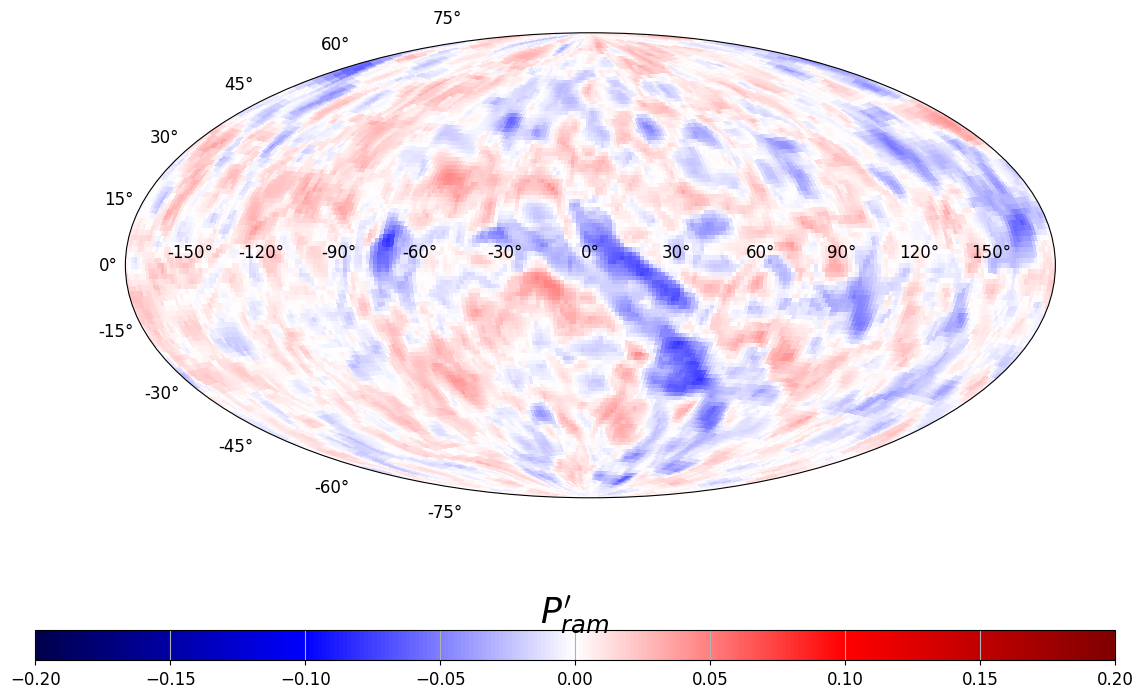}}

    \subfloat[Strong:0.09s]{\includegraphics[width=0.45\linewidth]{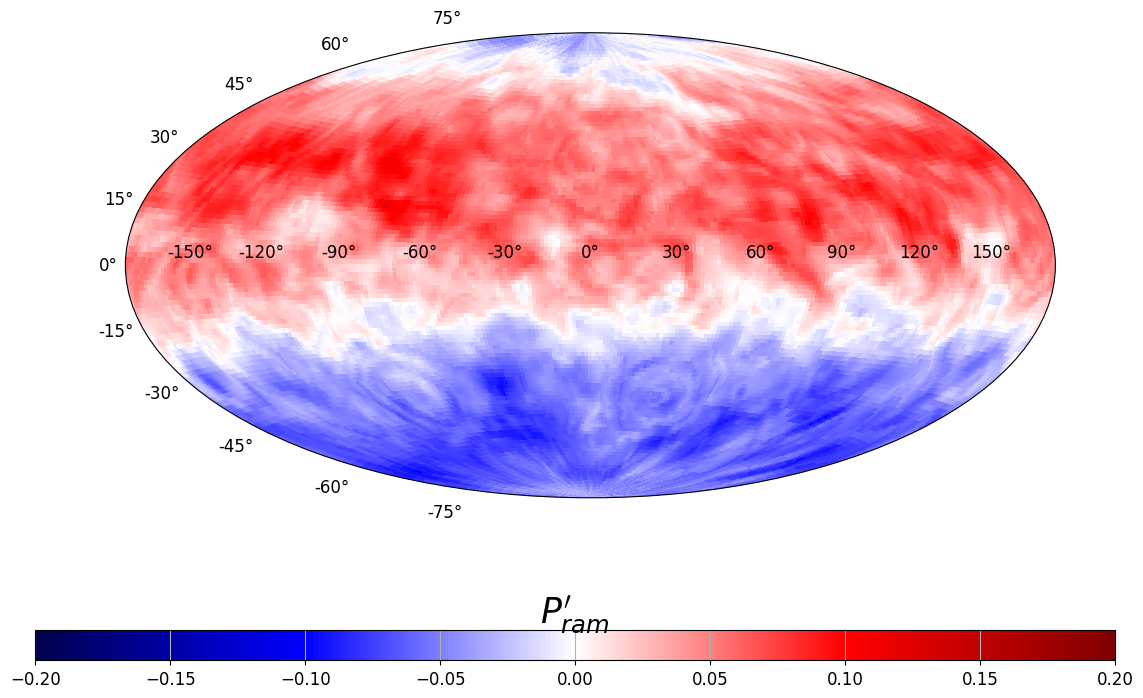}}
\hfil
    \subfloat[Weak:0.13s]{\includegraphics[width=0.45\linewidth]{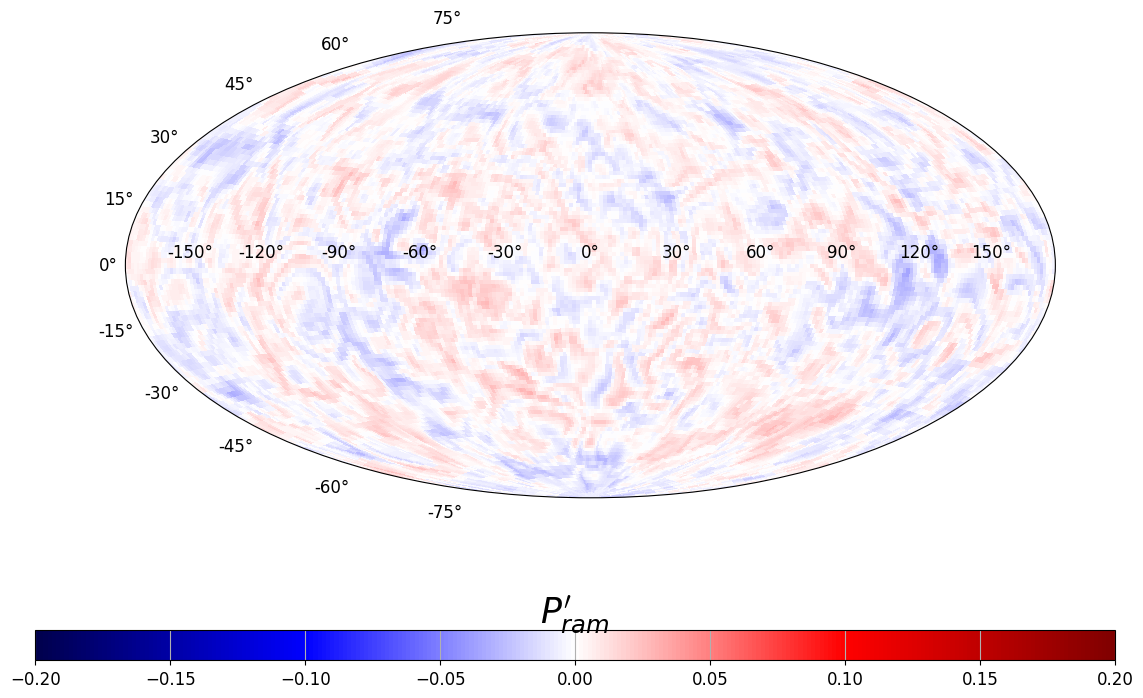}}
\caption{Relative ram pressure perturbations $P'$ ahead of the shock for the strong-field model (left column) and the weak-field model (right column)
at a radius of $200\, \mathrm{km}$ at different times (denoted below each panel). The final row shows the ram pressure perturbations amplitudes immediately before the maximum shock radius reaches $200\, \mathrm{km}$. }
    \label{fig:aitoff}
\end{figure*}

\begin{figure}
\centering
    \subfloat[(a)]{\includegraphics[width=\linewidth]{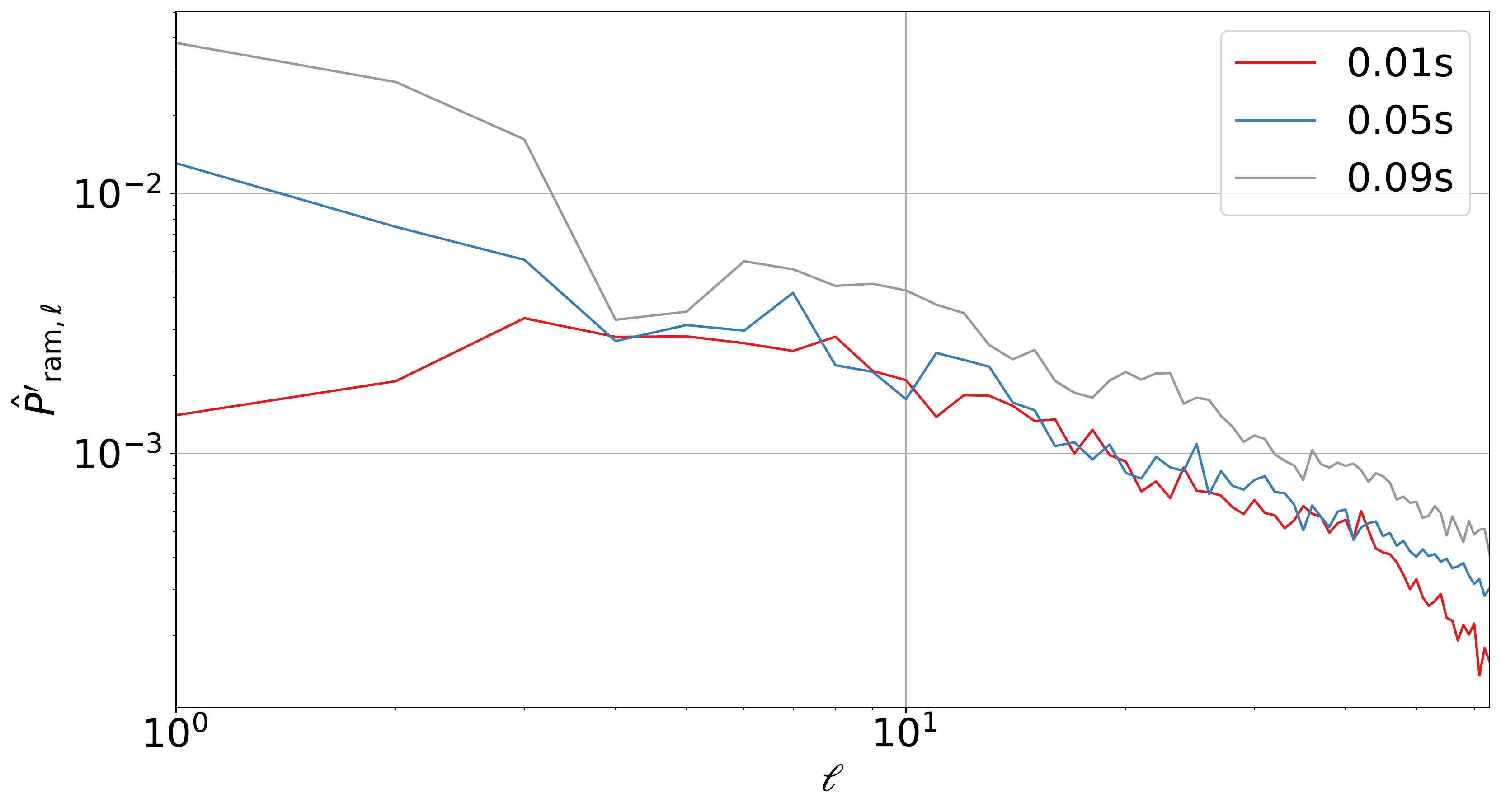}}
    
    \subfloat[(b)]{\includegraphics[width=\linewidth]{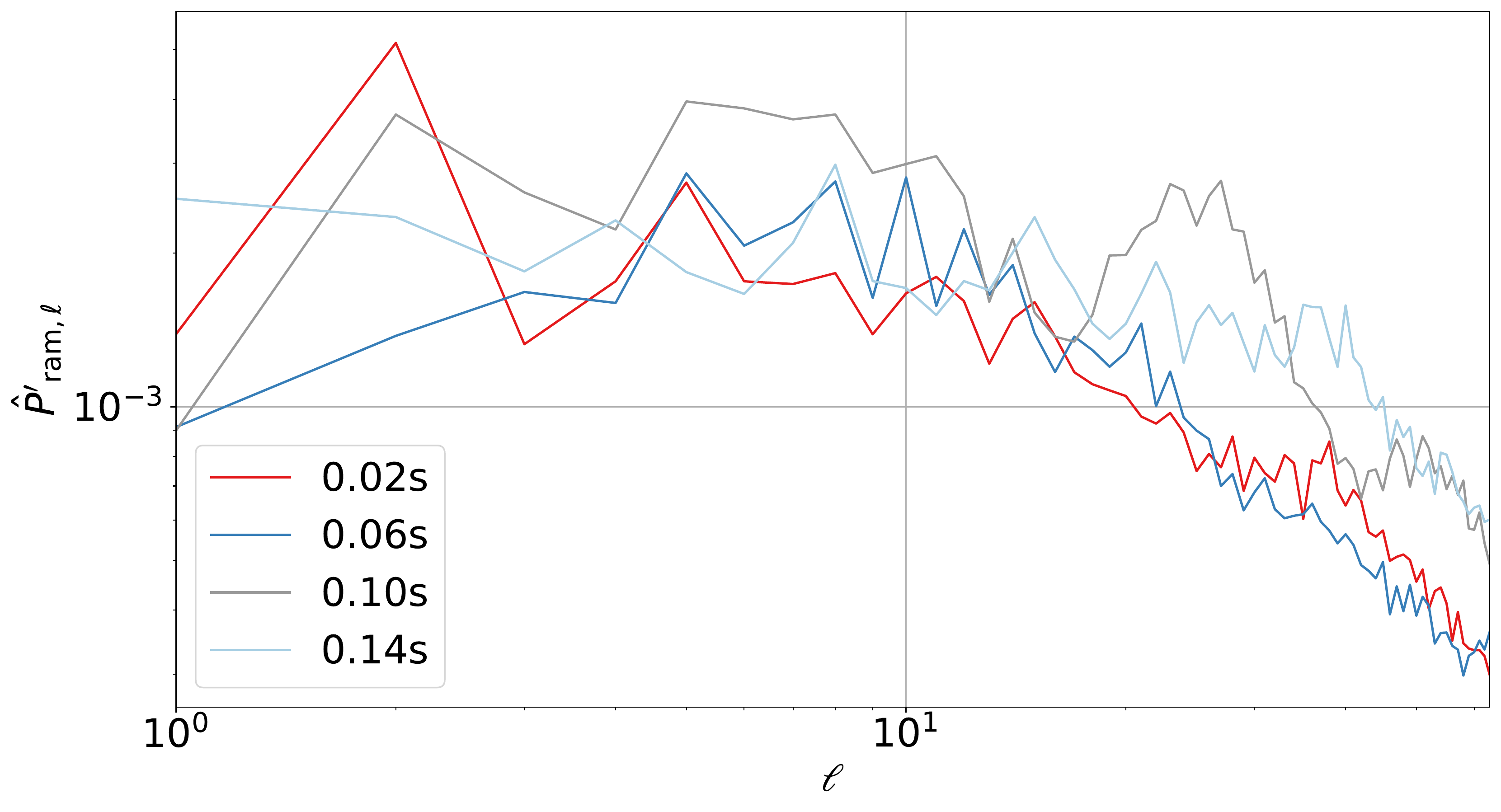}}

\caption{Spectrum $\hat{P}'_{\mathrm{ram},\ell}$ of the ram pressure perturbations 
(Equation~\ref{eq:specram})
at $200\, \mathrm{km}$, ahead of the shock, (a) for the strong-field model and (b) for the weak-field model.}
    \label{fig:Pram_spectra}
\end{figure}

\begin{figure*}
\centering
    \subfloat[Strong]{\includegraphics[width=0.45\linewidth]{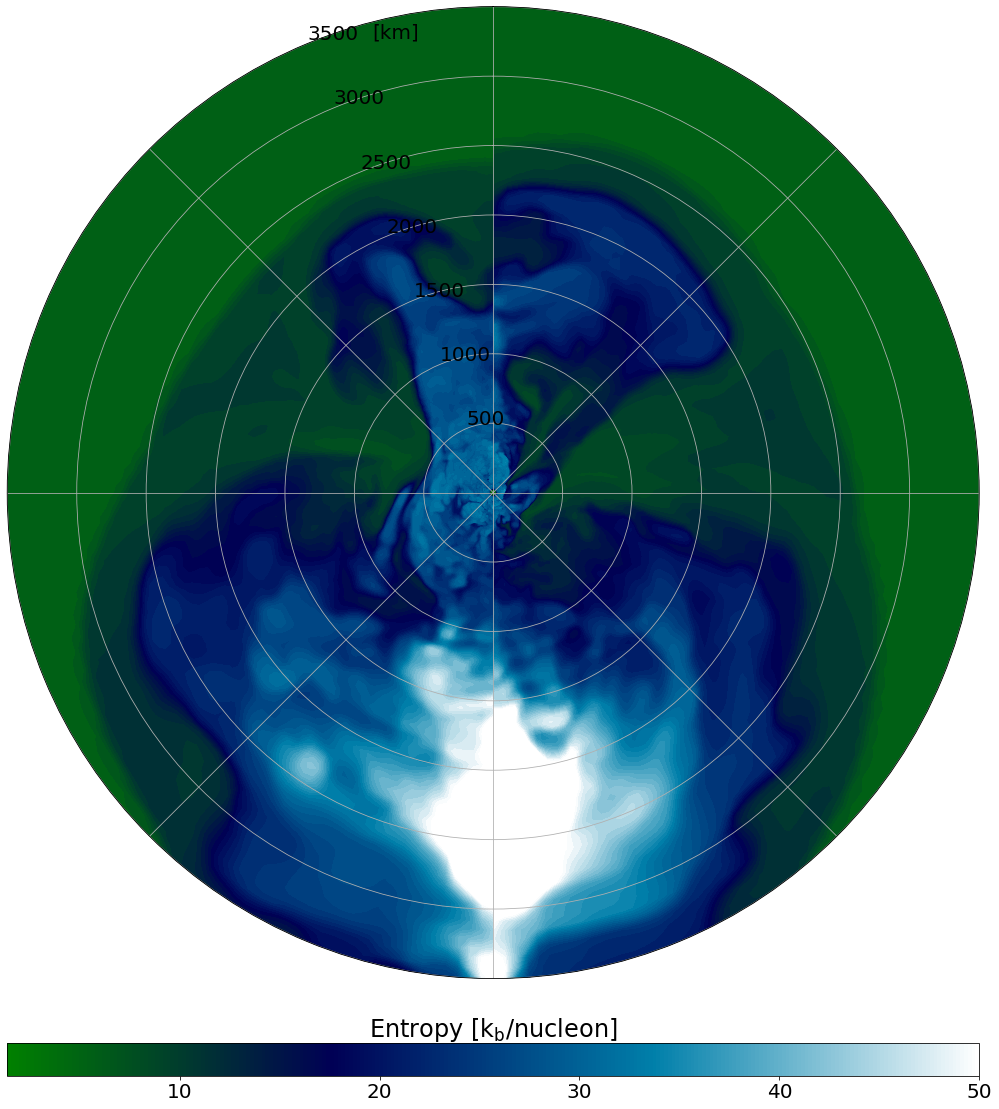}}
\hfil
    \subfloat[Weak]{\includegraphics[width=0.45\linewidth]{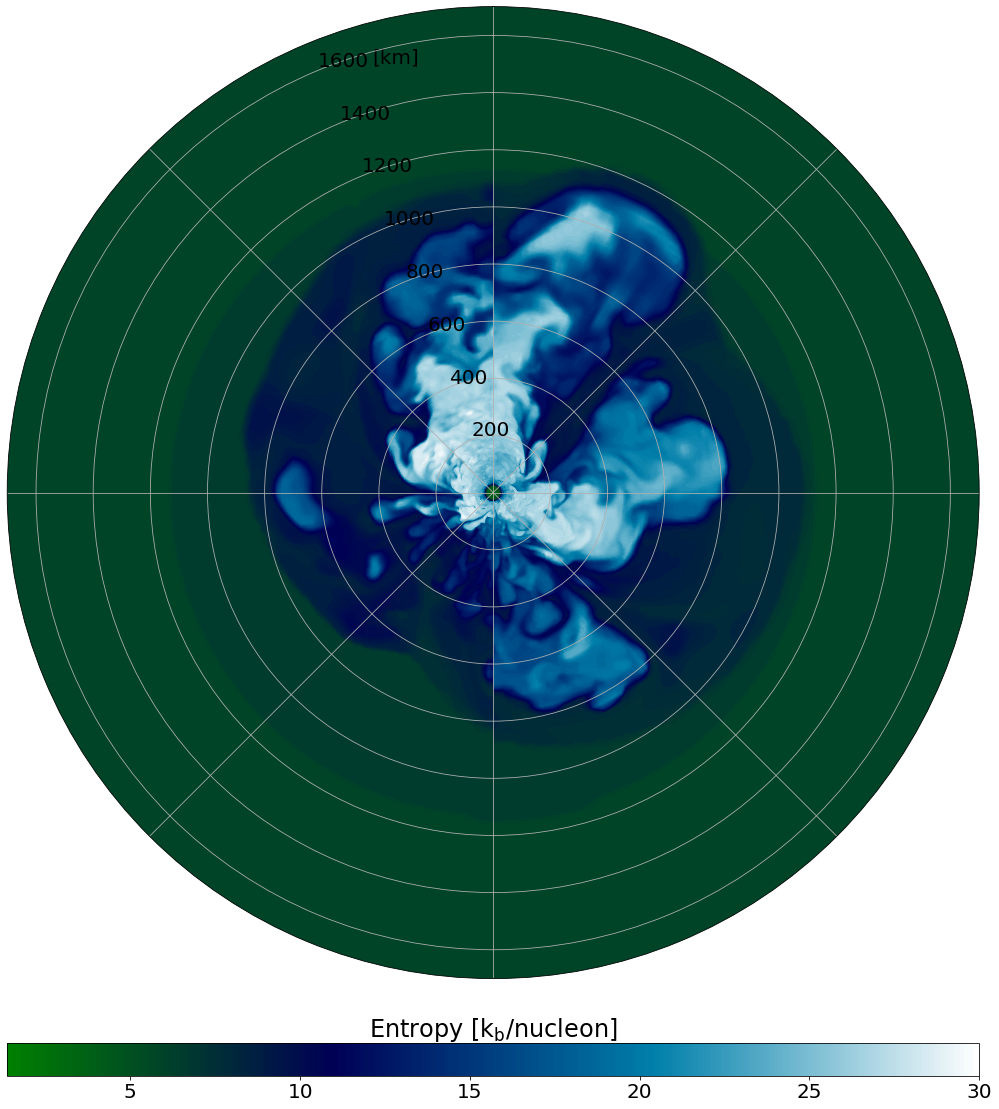}}

\caption{\Red{Entropy on meridional slices through the strong (left) and weak-field (right) model at $0.30\, \mathrm{s}$ after bounce. The strong-field model has a weakly bipolar explosion geometry with a prominent plume along the South polar axis and a weaker outflow along the North polar axis. The biggest neutrino-heated plume in the weak-field model is also aligned with the axis, but there are smaller plumes in other directions, most notably near the equatorial plane in the 4 o'clock direction. }}
    \label{fig:2D_Entropy}
\end{figure*}

In addition to increasing the turbulent in the gain region by jump-starting turbulent field amplification, the strong initial fields affect the dynamics by modifying the pre-shock conditions, similar to what occurs for ``perturbation-aided explosions'' in progenitors with vigorous convection in the oxygen or silicon shell \citep{Couch_2015,Muller2015b,Mueller_2017}. In the perturbation-aided mechanism, vortical convective motions are translated to sizable perturbations of the pre-shock density and ram pressure by advective-acoustic coupling
\citep{Muller2015b, abdikamalov_20,abdikamalov_21}, which then facilitate asymmetric shock expansion. 
We demonstrate that the initial fields assumed in our models also lead to differential infall and large-scale asymmetries in the pre-shock ram pressure.
To this end, we consider the relative ram pressure perturbations $P'_\mathrm{ram}$,
\begin{equation}
P' _\mathrm{ram} = \frac{P_\mathrm{ram} - \langle P_\mathrm{ram}\rangle}{\langle P_\mathrm{ram}\rangle},
\end{equation}
where $P_\mathrm{ram} = \rho v_r^2$ is the angle-dependent ram pressure, and  $\langle P_\mathrm{ram}\rangle$ is its average over solid angle.
We evaluate $P'_\mathrm{ram}$ at a radius of
$200\,\mathrm{km}$, which is small enough to capture the pre-shock conditions sufficiently well but larger than the maximum shock radius before shock revival.
2D Hammer projection of the ram pressure perturbations at this radius
are shown for the strong-and weak-field models at different times in Figure~\ref{fig:aitoff}.
In the strong-field model, large-scale asymmetries in the pre-shock ram pressure 
clearly emerge already at $0.05\, \mathrm{s}$
and grow considerably stronger shortly before
shock revival at $0.09\, \mathrm{s}$. The contrast between the directions of high and low ram pressure reaches about $20\%$, and there is a clear
\Red{large-scale asymmetry with a ram pressure deficit
in the Southern hemisphere, an excess at mid-latitudes in the Northern hemisphere, and
a deficit in a smaller region above latitudes
of $\gtrsim 60^\circ$ in the Northern hemisphere.} Such large-scale perturbations are most conducive to perturbation-aided shock revival \citep{Muller2015b}. The large-scale component of the perturbations is in fact more clearly visible than in simulations based on 3D progenitor models with oxygen shell convection
(cp.\ Figure~8 in \citealt{Mueller_2017}),
while the magnitude of relative pressure perturbation are of a similar magnitude.
In the weak-field model, the magnetic fields are not strong enough to dynamically affect the infall. Weaker pre-shock ram pressure perturbations emerge from numerical noise in the infall region in this case. These become weaker in time, and are dominated by small-scale structures.
Figure~\ref{fig:Pram_spectra} shows a spherical
harmonics decomposition of $P'_\mathrm{ram}$
to illustrate the different magnitude and scale of the ram pressure perturbations in the two models more quantitatively. We compute the spectrum $\hat{P}_{\mathrm{ram},\ell}$ as
\begin{equation}
\hat{P}_{\mathrm{ram},\ell}=
\sqrt{\sum_{m=-\ell}^\ell
\left|
\int Y_{\ell m}^\star P'_\mathrm{ram}\,\ud\Omega
\right|^2}.
\label{eq:specram}
\end{equation}
The spherical harmonic functions are orthonormalised and calculated using the SHTOOLS package \citep{wieczorek_16}.
The presence of a dipole \Red{and quadrupole} of growing amplitude
in the strong-field model is clearly
visible in Figure~\ref{fig:Pram_spectra}. \Red{The precise relation of the pre-shock ram pressure perturbations to the initial magnetic field structure is not yet fully transparent, and needs to be investigated further using linear perturbation theory in future (see discussion below). In particular, the emergence of an appreciable North-South asymmetry, which could be related to an instability of the initial twisted-torus field configuration, warrants further attention.}

\Red{As shown by Figure~\ref{fig:2D_Entropy}, the forcing by asymmetries in ram pressure is reflected in the explosion geometry of the strong-field model.
The most prominent high-entropy plume aligns with the region of lower ram pressure in the Southern hemisphere. There is a somewhat weaker neutrino-heated plume around the North polar axis as well, corresponding to the ram pressure deficit above latitudes of $60^\circ$ in the bottom left panel of Figure~\ref{fig:aitoff}. The prominent downflows at moderate latitudes in the Northern hemisphere are aligned with the regions of excess ram pressure. In the weakly magnetised model, the explosion geometry is characterised by a big plume along the North polar axis and several less prominent plumes. In this case, the emergence
of a prominent polar plume is likely facilitated by the grid geometry; a tendency towards grid-aligned explosion in weakly perturbed models has been observed before with the \textsc{CoCoNuT} code, though it is not universal (see, e.g., model he3.5 in \citealt{Mueller_2019}).
There is no clear relation to the pre-shock ram pressure perturbations. 
}

It should be noted, however, that the mechanism whereby magnetic fields are converted to acoustic perturbations during collapse still needs to be investigated in future work along the lines of existing linear theory for the amplification of hydrodynamic perturbations \citep{lai_00,takahashi_14,abdikamalov_20,abdikamalov_21}. 
\Red{The} nature of the magnetohydrodynamic perturbations
in the progenitor warrants further scrutiny. 
\Red{In the magnetohydrodynamic case, the linear
evolution during collapse can be expected to be
richer with interactions between between different
waves families beyond the mechanism of vortical-acoustic and entropic-acoustic coupling in the hydrodynamic case.}

\Red{There is also the question whether the
twisted-torus initial conditions adequately
approximate the magnetic field structure in
the progenitor.}
For the hydrodynamic case, it is important to take into account that the initial perturbation spectrum from
weakly compressible convective flow is dominated by vorticity and entropy perturbations and not by acoustic perturbations \citep{Muller2015b,abdikamalov_20}. In the absence
of consistent 3D progenitor models, this can adequately be captured by imposing a solenoidal flow geometry in convective regions in the progenitor.
In the fossil-field scenario, it is less clear whether imposing a reasonable twisted-torus field
geometry already captures the relevant physical constraints on the wave structure of the pre-collapse perturbations. 
\Red{This is particularly important for the strong-field case, where the magnetic fields
are strong enough to noticeably  perturb
hydrostatic equilibrium with initial values of a few $10^{-2}$ for the ratio
of magnetic to thermal pressure in the oxygen shell.}
For this reason, MHD models of quasi-equilibria with strong fossil fields in supernova progenitors are an important desideratum for the future.

\subsection{Neutron Star Properties}
\label{subsec:pns}
We next consider the birth properties of the neutron star produced in the two simulation. These are particularly relevant for evaluating the prospects for the fossil-field scenario for magnetar formation, although it must be borne in mind that the magnetic field and spin rate of the neutron star can still be affected by processes that occur on longer time scales than we can simulate here, such as spin-up and field burial due to fallback.

\begin{figure}
   \centering
  \includegraphics[width=\linewidth]{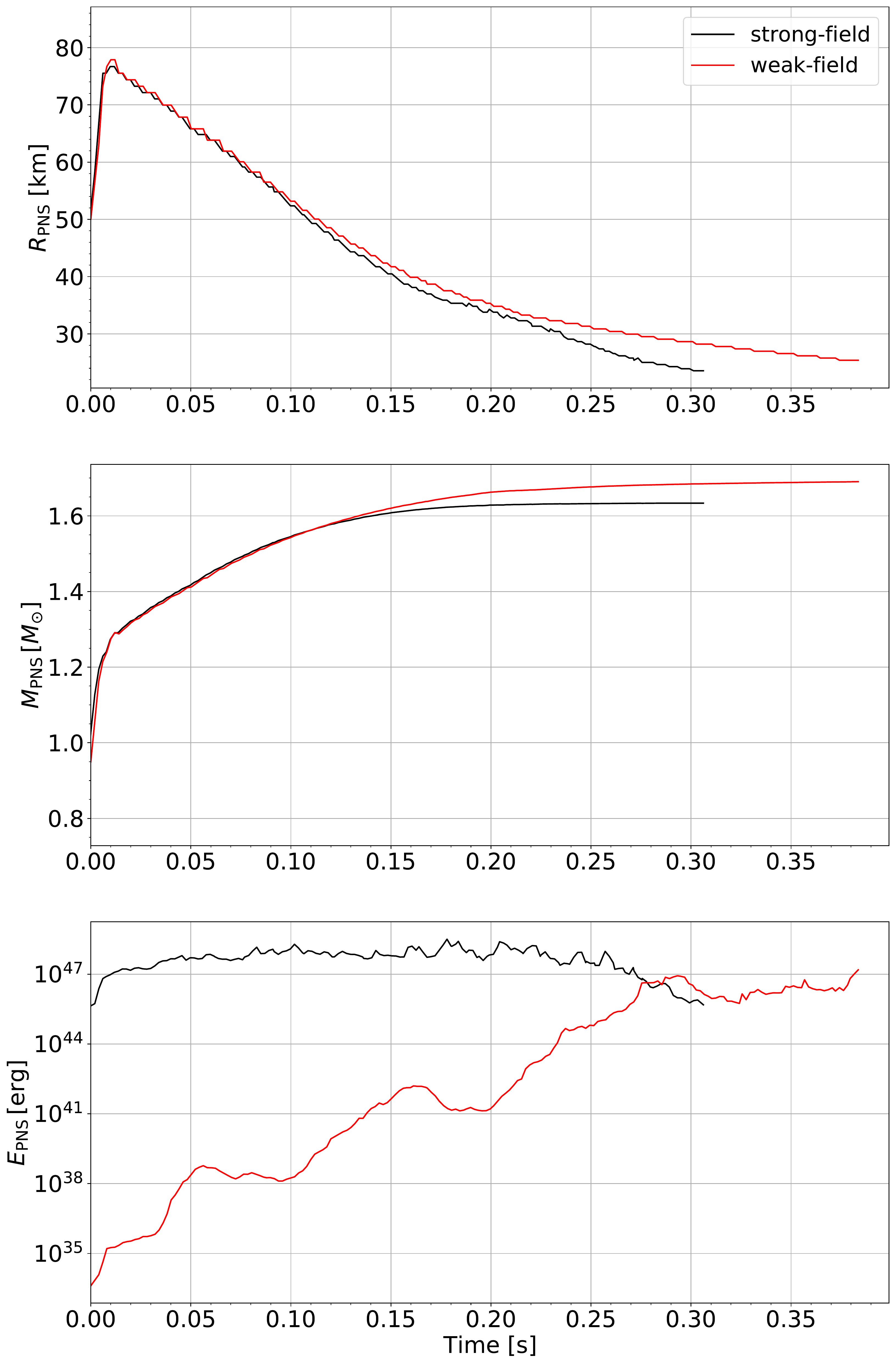}
  \caption{Evolution of the radius $R_\mathrm{PNS}$, baryonic mass $M_\mathrm{PNS}$, and magnetic energy  $E_\mathrm{PNS}$ of the PNS for the strong-field model (black) and the weak-field model (red). The PNS surface is defined as the radius where the angle-averaged density reaches $10^{11}\,\mathrm{g}\,\mathrm{cm}^{-3}$.
  }
  \label{fig:PNS}
\end{figure}

We show the bulk properties of the PNS in Figure~\ref{fig:PNS}, namely is radius ${R_\mathrm{PNS}}$, mass ${M_\mathrm{PNS}}$, and magnetic energy ${E_\mathrm{PNS}}$, which we compute as volume integrals over the region where $\rho > 10^{11} \mathrm{g \,cm^{-3}}$. Initially, the PNS mass and radius for both models are largely identical until $\mathord{\approx} 0.15\,\mathrm{s}$ aside from minimal differences resulting from small differences during the collapse phase. After this point, $M_\mathrm{PNS}$
continues to grow in the weak-field model and
the model maintains a larger $R_\mathrm{PNS}$ since accretion continues longer than in the strong-field model.  The final baryonic masses are
$1.63\, M_\odot$
and $1.69\, M_\odot$ for the strong- and weak-field model, respectively. Using the approximation of \citet{Lattimer1989, Lattimer2001} for the neutron star binding energy, the
gravitational mass $M_\mathrm{grav}$
can be obtained from
\begin{equation}
    M_\mathrm{by} = M_\mathrm{grav} + 0.084M_{\odot}(M_\mathrm{grav}/M_{\odot})^2,
\end{equation}
which results in $M_\mathrm{grav}=\,1.46M_{\odot}$ 
for the strong-field model
and $M_\mathrm{grav}=\,1.50M_{\odot}$
for the weak-field model.

The evolution of the interior and surface field of the PNS is of particular interest in the context of magnetar and pulsar birth scenarios.
Due to the vast differences in initial magnetic field strength, the initial $E_\mathrm{PNS}$
(bottom panel of Figure~\ref{fig:PNS}) differs by orders of magnitude between the strong-and weak-field models. For the strong-field model $E_\mathrm{PNS}$ remains rather steady at  $10^{47 \texttt{-} 48}\, \mathrm{erg}$ until about $0.25\, \mathrm{s}$ after bounce.  In the weak-field model, $E_\mathrm{PNS}$ starts at much smaller values, but gradually grows due to compression of the field lines as the 
$\mathrm{R_{PNS}}$ contracts, shear flows in the PNS surface region, and amplification by PNS convection. Eventually $E_\mathrm{PNS}$
becomes comparable to the strong-field model about  $0.25\, \mathrm{s}$ after bounce.
It is noteworthy that by the end of the simulation, the magnetic energy in the PNS is dropping in the strong-field model and ends up slightly lower than in the weak-field model. The drop in magnetic energy for the strong-field model is likely due to numerical reconnection and variations in field amplification by PNS convection.
While turbulent reconnection is bound to occur in the PNS convection zone, the destruction of magnetic fields may be overestimated numerically, especially at the modest resolution affordable inside the PNS. Nonetheless, the simulations at least strongly
indicate that the memory of the initial fields is largely lost inside the PNS.\footnote{PNS fields originating from field with significant helicity might still be protected better from field decay in the long run, but once PNS convection has died down and differential rotation has been eliminated, field decay will be extremely slow because of the high conductivity of neutron star matter, regardless of the helicity of the field configuration.}

\begin{figure}
\centering
    \subfloat[(a)]{\includegraphics[width=\linewidth]{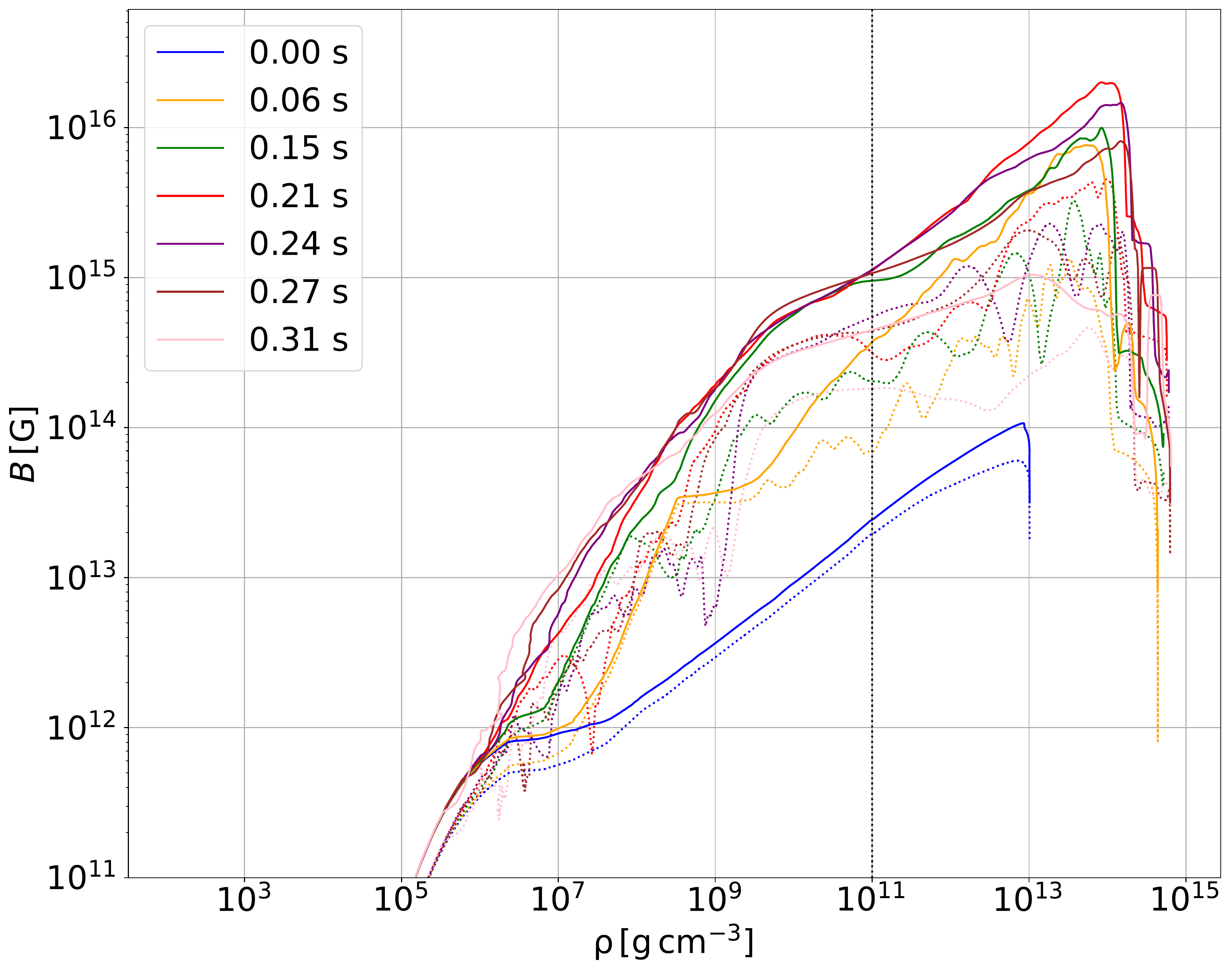}}

    \subfloat[(b)]{\includegraphics[width=\linewidth]{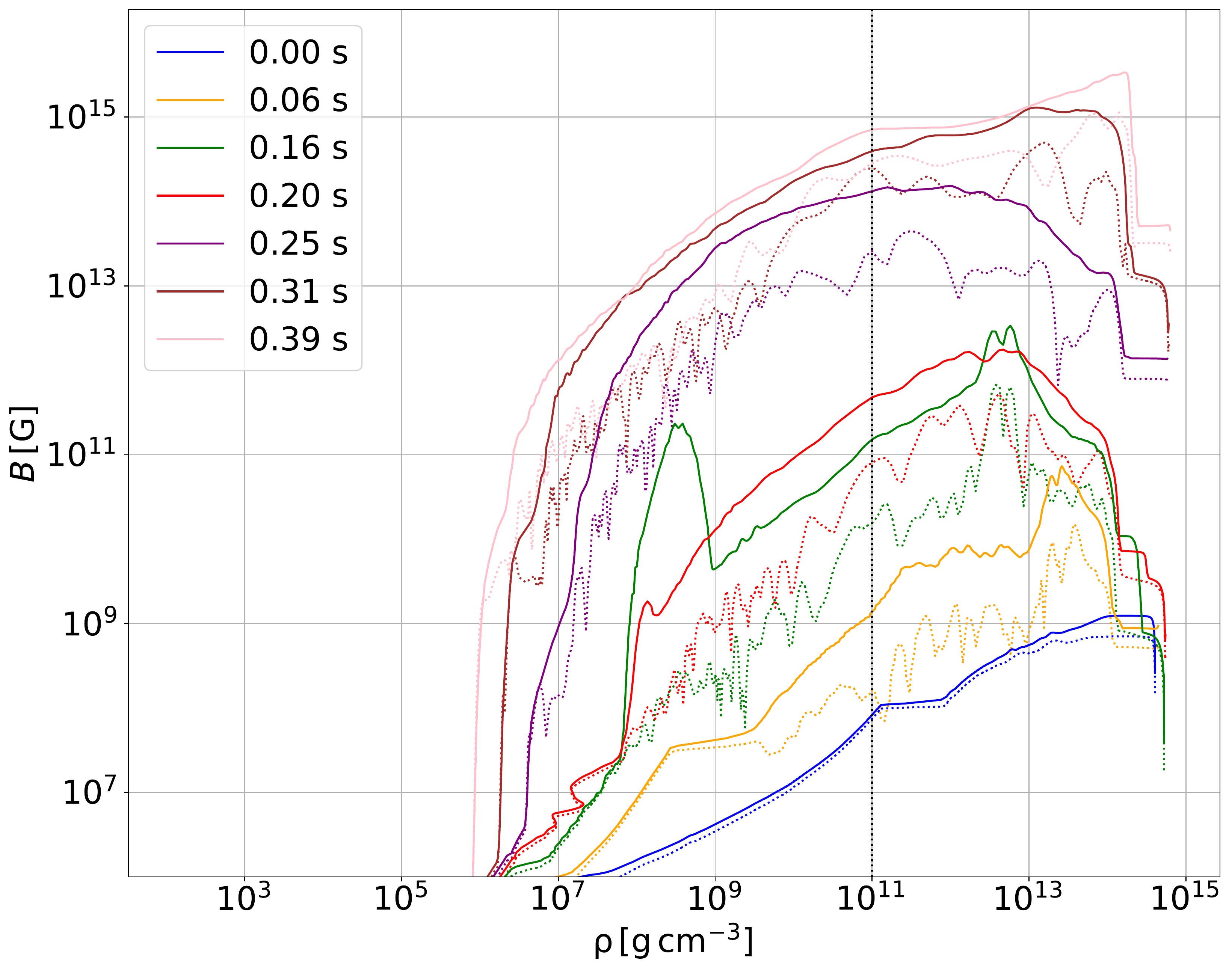}}

\caption{Angle-averaged total magnetic field strength (solid lines, root-mean-square averages), and the dipole component of the
radial magnetic field $B_r$ (dotted lines) 
as a function of density (a) for the strong-field model and (b) for the weak-field model. The vertical dotted black lines indicate 
the fiducial PNS surface density
$\rho = 10^{11}\mathrm{g\,cm^{-3}}$. }
    \label{fig:Brho_profile}
\end{figure}

To elucidate the spatial structure of the PNS
magnetic field,
we plot profiles of the
dipole strength of the radial magnetic field
and the root-mean-squared average of the total field strength against density for the strong- and weak-field models in  Figure~ \ref{fig:Brho_profile}.
The tentative ``surface'' of the PNS where $\rho = 10^{11}\,\mathrm{g}\,\mathrm{cm}^{-3}$ is indicated by dotted vertical line, but it must be borne in mind that further accretion, ablation of surface material by the neutrino-driven wind, and possibly field breakout from the PNS may yet change the PNS surface structure and surface magnetic fields in particular. Just as for the bulk magnetic energy of the PNS, the PNS surface magnetic
field has effectively lost  memory of the initial condition after less than $0.3\,\mathrm{s}$ after bounce, even though the initial magnetic field strengths of both these models were six orders of magnitude apart.
At the final simulation time, the surface magnetic field strengths in both models are very similar, with dipole fields of
$5\times10^{14}\,\mathrm{G}$ for the weak-field case and $2\times10^{14}\,\mathrm{G}$
for the strong field case. Thus, both models reach magnetar-strength fields at the surface as in previous 3D MHD supernova simulations of non-rotating and slowly-rotating progenitors
\citep{MullerVarma2020,Matsumoto2022}, at least transiently. It is interesting, however, that the surface field in the strong-field model is decreasing at the end of the simulation, which may reduce possible tensions with the observed field strength distribution of young neutron stars. The declining field strength is consistent with the development of net outflow
from the PNS surface, i.e., an incipient neutrino-driven wind, which implies a change from flux compression to flux expansion. The final surface field strength is obviously still beyond the scope of the current simulations.

\begin{figure}
   \centering
  \includegraphics[width=\linewidth]{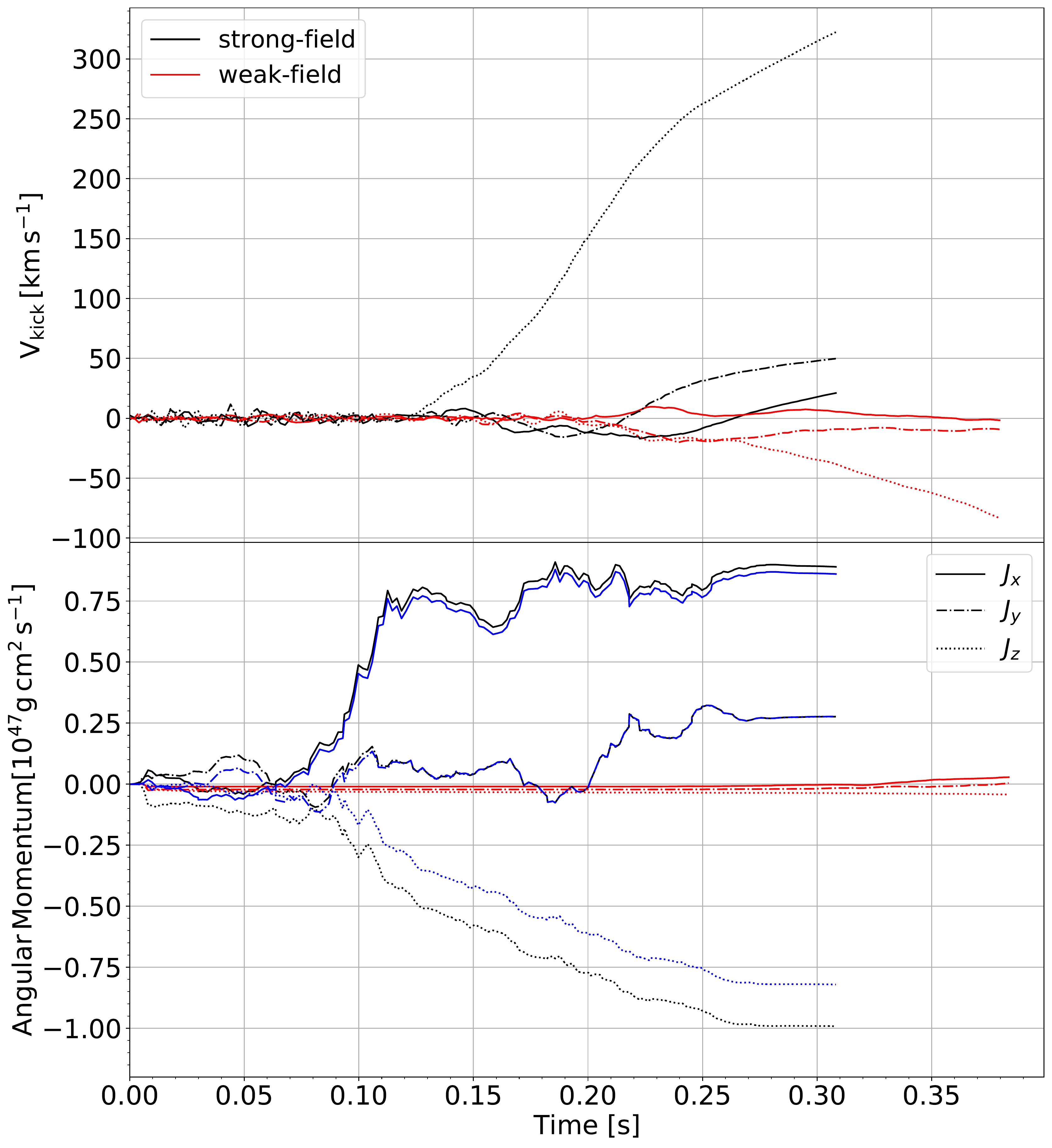}
  \caption{Evolution of the kick and angular momentum of PNS for
  the strong-field model (black/blue) and the weak-field model (red).
  For the strong-field model, we show both the time integral of the
  entire angular momentum flux onto the PNS (black), and the
  time integral of magnetic torques alone (blue). The advective
  angular momentum flux and the magnetic torques are evaluated
  on a sphere of radius $30\,\mathrm{km}$ around the origin. 
  Different line styles are used for the three components of
  the angular momentum vector (solid: $J_x$, dash-dotted: $J_y$, dotted: $J_z$).
  }
  \label{fig:PNS_kick}
\end{figure}

We next consider the kick velocity and angular momentum of the newly born neutron star, which we plot in Figure~\ref{fig:PNS_kick}. 
Following \citet{Scheck2008}, we evaluate the kick indirectly from the momentum of the ejecta assuming angular momentum conservation since the innermost part of our grid is treated in spherical symmetry and does not allow the PNS to move freely. Specifically, we compute
\begin{equation}
    \mathbf{v}_\mathrm{kick} = - \frac{1}{M} \int_{r>R_\mathrm{gain}} \rho \mathbf{v} \,\ud V,
\end{equation}
where $M$ is the total mass\footnote{Although the choice between the baryonic and gravitational mass would only make a minor difference here, we note that it is appropriate to use the baryonic PNS mass for evaluating the kick in a pseudo-Newtonian simulation. 
Although the inertial mass of the neutron star is eventually reduced by neutrino emission as it cools,
near-isotropic emission of neutrinos in the PNS rest frame will not change the PNS velocity, and decrease the PNS momentum proportionately to the decrease in inertial mass.} inside the gain radius, and the integral covers the entire volume outside the gain radius. 
\Red{In both models, the kick is almost aligned with the $z$-axis of the spherical polar grid, reflecting the explosion geometry shown in Figure~\ref{fig:2D_Entropy}, and thus
the forcing by initial perturbations and/or
the spherical polar grid geometry (cp.\ discussion in Section~\ref{subsec:gaindynamics}). As expected, the kick points in the direction opposite to the
most prominent plume, i.e., towards the North
pole in the strong-field model and towards the
South pole in the weak-field model, at least
by the end of the simulation.}
Since the explosion in the weak-field model has not yet sufficiently advanced to extrapolate the final value of the kick, which has only reached $84\, \mathrm{km}\,\mathrm{s}^{-1}$ by the end of the simulation. \Red{The kick direction may also
still be subject to change, since other non-grid
aligned plumes and downflows are still present
in the weak-field model and may impart momentum
onto the PNS in other directions.}
In the strong-field model, the kick reaches $326\, \mathrm{km}\, \mathrm{s}^{-1}$, and although it is still growing, its rate of change is decreasing. Since accretion downflows have already been quenched and the ejecta are expanding steadily, the final kick will only grow further by a modest amount. We can estimate the final kick velocity of these models using the extrapolation method demonstrated in \cite{Mueller_2019} to find final PNS kick velocities of $345\, \mathrm{km}\, \mathrm{s}^{-1}$ and $110\, \mathrm{km}\, \mathrm{s}^{-1}$ for the strong and weak-field simulations, respectively.
We thus expect the final kick to lie within the
typical range observed for young neutron stars
\citep{hobbs_05}.

The second panel in Figure~\ref{fig:PNS_kick} shows the angular momentum $\mathbf{J}$ of the PNS, which is obtained by integrating
the magnetic torques and hydrodynamic angular momentum fluxes onto the PNS in time and over the surface of a sphere of radius $30\, \mathrm{km}$,
\begin{equation}
    \frac{\ud \mathbf{J}}{\ud t}
    =-\oint (\mathbf{r}\times  \mathbf{v}) \rho v_r\,\ud A
    +\frac{1}{4\pi}\oint (\mathbf{r}\times \mathbf{B}) B_r\,\ud A.
\end{equation}
For the strong-field model, the contribution from magnetic torques is also shown separately. We note, however, that our choice of the PNS surface being at $30\, \mathrm{km}$ is somewhat arbitrary; other works such as \citet{wongwathanarat_13} have used a considerably larger analysis region up to radii of several hundred kilometres. The choice of a small radius is designed to better capture the current PNS spin,
but the final spin of the PNS depends rather sensitively on how much of the material in its vicinity is eventually accreted or blown out\footnote{The sensitivity of the PNS spin to rather small amounts of accretion has been demonstrated most perspicuously in the extreme case of fallback accretion at later times, which can easily spin up the PNS to millisecond periods \citep{Chan2018,Chan2020,Stockinger2020}.}. The eventual multi-dimensional ``mass cut'' cannot be predicted at this rather early stage of the explosion yet. This introduces a significant uncertainty on the final spin, since the shells outside $30\, \mathrm{km}$ still carry considerable angular momentum and have not yet been brought into rigid rotation with the PNS by the end of the simulation. Experiments with different analysis radii (corresponding to different hypothetical mass cuts) indicate that the PNS angular momentum could still change by a small factor, with the direction of the spin appearing more uncertain than the spin magnitude. By contrast, $\mathbf{v}_\mathrm{kick}$ is quite insensitive to the precise choice of the analysis region. The results for the weak-field model are only shown for the sake of completeness; as for the kick, the angular momentum shows no sign of saturation in this case yet. Its absolute value at the end of the simulation is $5.11 \times 10^{45}\,\mathrm{g}\,\mathrm{cm}^2\,\mathrm{s}^{-1}$ and is still growing.

In the strong-field model, the angular momentum of the PNS has largely frozen out, however. 
By the end of the simulation, it has reached about $1.36 \times 10^{47}\,\mathrm{g}\,\mathrm{cm}^2\,\mathrm{s}^{-1}$.  It is interesting to note that the spin-up of the PNS in the strong-field model is driven mostly by magnetic torques.

\begin{figure}
   \centering
  \includegraphics[width=\linewidth]{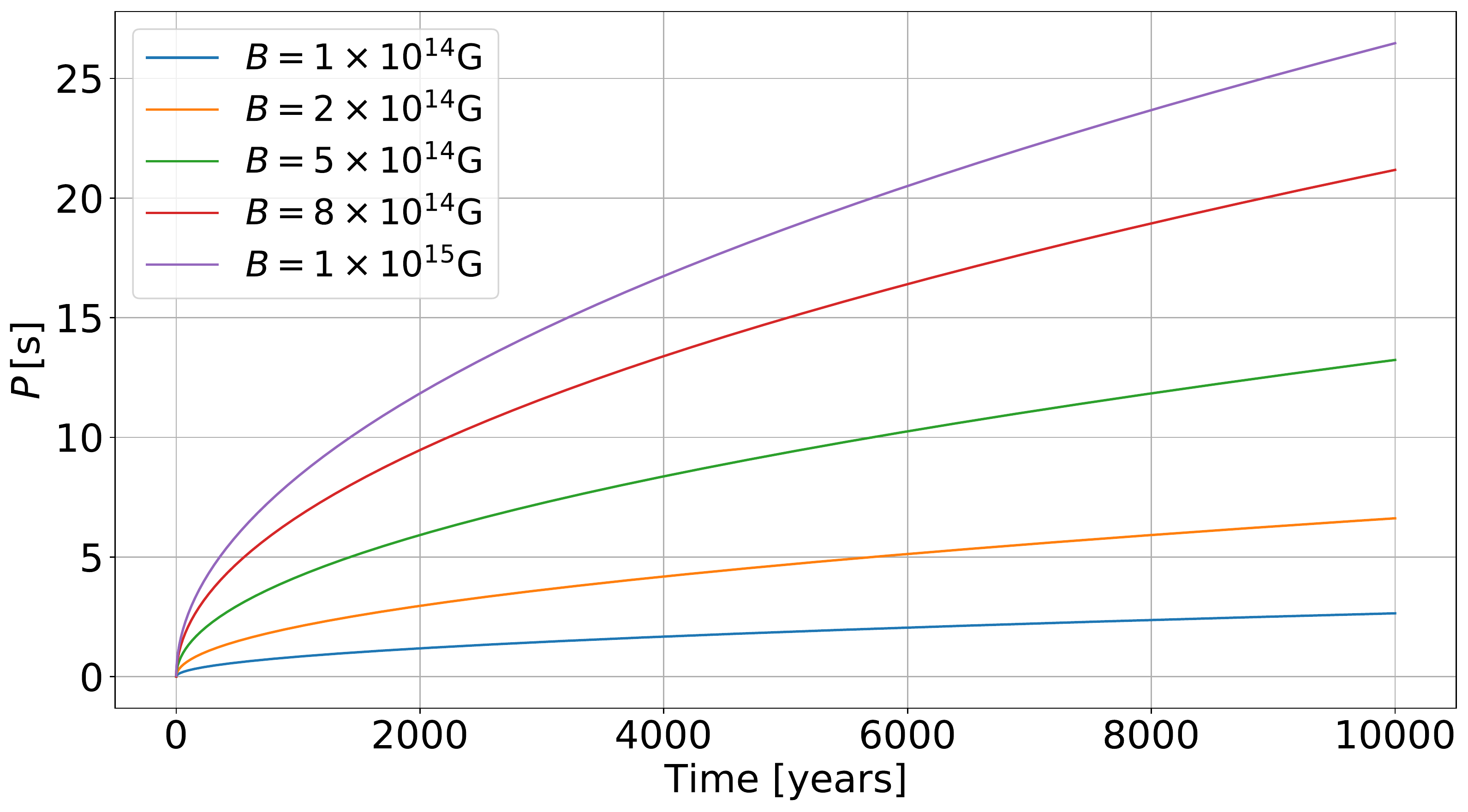}
  \caption{Rotation period of the neutron star as a function
  of age using the spin-down formula (\ref{eq:spindown}).}
  \label{fig:period}
\end{figure}

Using the analytic approximation for the neutron star moment of inertia from \citet{Lattimer2005},
\begin{equation}
    I \approx 0.237 M_\mathrm{grav}R^2
    \left[
    1 + 4.2\left(\frac{M_\mathrm{grav}}{M_{\odot}}\frac{\mathrm{km}}{R}\right)
    + 90\left(\frac{M_\mathrm{grav}}{M_{\odot}}\frac{\mathrm{km}}{R}\right)
    \right],
\end{equation}
the angular momentum can be translated into a neutron star spin period $P=J/I$. Assuming a final neutron star radius $R$ of $12\, \mathrm{km}$, we estimate the moment of inertia as  $\mathrm{1.52 \times 10^{45}g\,cm^3}$  for the strong-field model
and  $\mathrm{1.58 \times 10^{45}g\,cm^3}$ for the weak-field model, which corresponds to spin periods of  about $73\, \mathrm{ms}$ and $1.86\, \mathrm{s}$, respectively. Prima facie, this puts the strong-field model in the range of typical spin-period of young \emph{pulsar} \citep{faucher_06,perna_08,popov_12,igoshev_13}, and not of magnetars, which are observed to rotate rather slowly \citep{Olausen2014,kaspi_17}.
However, highly magnetised neutron stars will also spin-down rapidly; even the youngest observed magnetars must already have been spun down considerably after the supernovae.
We therefore extrapolate the spin period using
the neutron star spin-down formula for the period derivative $\dot{P}$ in terms of $P$ and the surface dipole field $B_\mathrm{surf}$
 \citep[cp.\ ][]{Lorimer2004},

\Red{
\begin{equation}
    \dot{P} = \frac{1}{P} \left(\frac{B_\mathrm{surf}}{3\times 10^{19}}\right)^{2} \,\mathrm{s^{-1}}.
    \label{eq:spindown}
\end{equation}
}
Since the final surface dipole field strength is
still subject to uncertainties, we calculate the spin-down for several choices of $B_\mathrm{surf} $
in the range around the tentative dipole field strength
of $2\times 10^{14}\,\mathrm{G}$ at the end of our simulation.
The extrapolated spin period after several hundred years for the strong-field model (Figure \ref{fig:period}) reach several seconds, which is compatible with observations \citep{Olausen2014}, which show magnetar spin periods of $1\, \mathrm{s} < P < 12\, \mathrm{s}$. The results can, e.g., be compared to Swift J1818.0–1607 \citep{Blumer2020},
one of the youngest known magnetars, which is approximately 500 years old and has a spin period of $1.36\,\mathrm{s}$. For a surface dipole  field strength of $10^{14}\,\mathrm{G}$, we obtain a spin period of $\mathord{\sim}0.6\,\mathrm{s}$ after 500 years. While this is still faster than Swift J1818.0–1607, it is of the same magnitude, and thus appears roughly compatible with observed constraints on magnetar spins, especially considering that stochastic variations among strongly magnetised neutron stars are expected.  

\begin{figure}
   \centering
  \includegraphics[width=\linewidth]{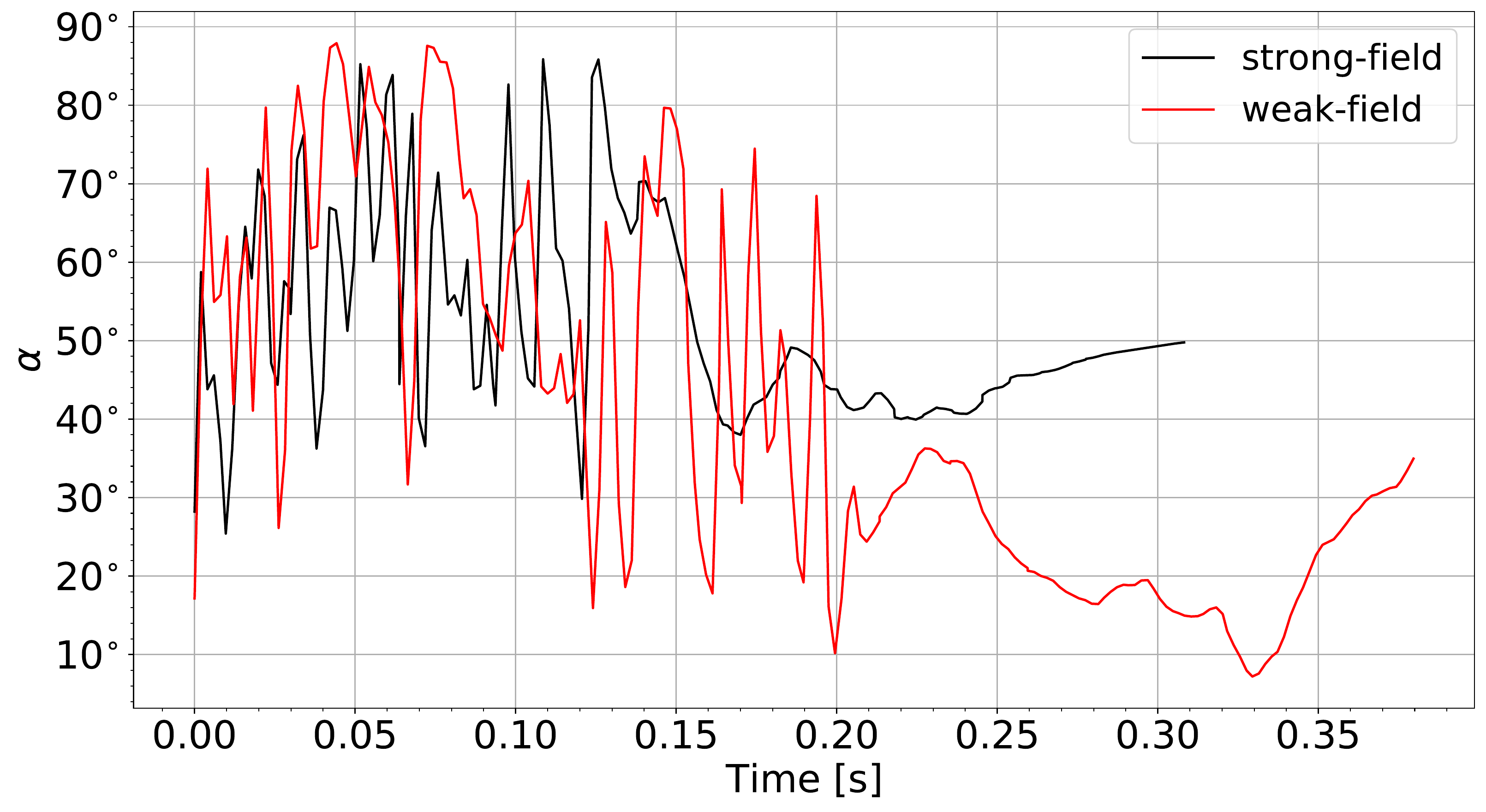}
  \caption{The angle $\alpha$ between the spin and kick directions of the PNS based on an evaluation of the
  PNS angular momentum using the hydrodynamic fluxes
  and magnetic torques at a radius of $30\,\mathrm{km}$.}
  \label{fig:alpha}
\end{figure}

Finally, we consider the orientation of the kick
and spin direction. Observational evidence has suggested generic spin-kick alignment in neutron stars \citep{Johnston2005, Noutsos2012, Noutsos2013},
but simulations have yet to reveal a robust mechanism for spin-kick alignment; recent attempts to connect spin-kick alignment to rotation \citep{powell_20} or fallback \citep{Janka2022} have not yielded wholly convincing results. At least during the early explosion phase, magnetic fields in non-rotating
supernova models do not appear to provide a mechanism
for spin-kick alignment either, as can
be seen from Figure~\ref{fig:alpha}, which shows the angle $\alpha$ between the spin and kick direction for our models. We find a misalignment of about $ 50^\circ$  and $35^\circ$ at the end of the simulation for the strong-and weak-field models. 
With supernova explosion models now including magnetic fields, it becomes increasingly difficult to account for spin-kick alignment during the early explosion phase. Spin-kick alignment may indeed be a long-term
effect and related, e.g., to later fallback \citep{Janka2022} (although the mechanism proposed by \citealt{Janka2022} likely does not work as intended because of a problem with angular momentum non-conservation).

\section{Conclusions}
\label{sec:conclusion}
We investigated the role of strong magnetic fields in non-rotating progenitors on core-collapse supernova explosions, specifically its effect on shock revival, explosion dynamics, and the properties of the compact remnant. To this end, we performed 3D simulations of the collapse and explosion of a $16.9 M_\odot$ star using a twisted-torus initial field and two different initial field strengths ($10^{12}\, \mathrm{G}$ and $10^{6}\, \mathrm{G}$ at the centre). It has been speculated that strong pre-supernova fields of order $10^{12}\, \mathrm{G}$ might arise in massive stars as fossil fields from a stellar merger \citep{ferrario_05,ferrario_06,SchneiderNature,Schneider2020}.

Our simulations show that strong magnetic fields have a noticeable impact on the neutrino-driven explosion mechanism, even in slowly/non-rotating progenitors.
The strong-field model undergoes  shock revival
already early on about $0.1\,\mathrm{s}$ after bounce
and reaches an explosion energy $9.3 \times 10^{50}\, \mathrm{erg}$ by the end of our simulation
at $0.3\, \mathrm{s}$ after bounce. Thus, the model
reaches a typical supernova explosion energy faster than in non-magnetised 3D models of core-collapse supernovae from slowly rotating progenitors.
While the corresponding weak-field model also explodes somewhat later, it shows a much slower rise of the explosion energy to only $1.2 \times 10^{50}\, \mathrm{erg}$ by the end of the simulation.

The principal role of magnetic fields in the precipitous and powerful explosion of the strong-field model is to provide an extra boost to neutrino heating and hydrodynamic turbulence to trigger the explosion early, while the explosion is still primarily driven by neutrinos. This is similar to the results of \citet{MullerVarma2020}, who already found that field amplification by a turbulent dynamo can aid shock revival by the neutrino-driven mechanism; the key difference is that the effect of magnetic fields comes into play early on without the need for an extended growth phase. The saturation field strength in the gain region, corresponding to about 40\% of kinetic equipartition
is reached almost immediately once neutrino-driven convection develops. In the weak-field model, field amplification by the turbulent dynamo is too slow to have a substantial impact before a neutrino-driven explosion develops

As in \citet{MullerVarma2020}, magnetic fields aid shock revival in the strong-field model by increasing the convective efficiency in the gain region and reducing the binding energy of the gain region. In addition, the strong fields in the progenitor also create sizable perturbations 
of $\pm 10\%$ in the pre-shock ram pressure, reminiscent of the effect of infalling convective eddies from shell burning in perturbation-aided explosions \citep{Muller2015b,Mueller_2017,abdikamalov_20}.
The perturbations are dominated by a large-scale $\ell=1$ dipole mode. Large-scale ram-pressure perturbations of such amplitudes tend to be strongly
conducive to shock revival \citep{Muller2015b,abdikamalov_16,huete_18}.
In future, the evolution of strong magnetic seed fields
in the progenitor and their impact on shock revival need to be investigated using a more consistent approach by basing supernova simulations directly
on models of magnetoconvection in massive stars \citep{Varma2021} and complementing simulations
with analytic theory for the amplification and
dynamical role of seed perturbations, similar to
the hydrodynamic case \citep{takahashi_14,Mueller_2016,huete_18,abdikamalov_20,abdikamalov_21}.

The high explosion energy of the strong-field model is primarily a consequence of early shock revival. Its energetics is consistent with the physics known from other models of neutrino-driven explosions \citep{Marek2006,Muller2015a,Bruenn2016,Mueller_2017}. Nucleon recombination is primarily, but not exclusively, responsible  for the delivery of the explosion energy. Due to early shock revival, the volume-integrated neutrino heating rate is high and the specific binding energy of the gain region is modest. These factors allow for a high mass outflow rate, a very efficient conversion of neutrino heating into recombination energy in the ejecta region, and hence a rapid growth of the explosion energy. As in early explosions in 2D simulations \citep{Bruenn2016}, mechanical  $P\,\ud V$ work and (specific to  our model) magnetic stresses also contribute significantly to the energy budget of the ejecta.

The strong-field model has evolved sufficiently far to tentatively extrapolate some properties of the neutron star born in the supernova. Our simulation predicts a gravitational neutron star mass of  $1.50 M_\odot$, a
kick velocity of about $345\,\mathrm{km}\,\mathrm{s}^{-1}$, and a birth spin period of about $73\, \mathrm{ms}$. The dipole component of the magnetic field at the surface of the proto-neutron star is about $2\times 10^{14}\,\mathrm{G}$ at the end of the simulation, but is decreasing significantly at this point. Such a field could spin down the neutron star to periods of second within a few hundred years, which would be compatible with the spin period of young magnetars. We do not find spin-kick alignment, and the mechanism that might bring about such an alignment remains opaque.

However, while the raw neutron star parameters from the simulation appear to be quite consistent with observational constraints on young neutron stars and magnetars in particular, our simulations rather prompt more questions about neutron star formation channels and birth properties. It remains unclear whether the proto-neutron star in the strong-field would actually evolve into a magnetar. The surface dipole field strength is decreasing by the end of the simulation, and we cannot follow phenomena that may shape its final magnetic fields on much longer time scales, such as
the ablation of surface material by the neutrino-driven wind, field burial \citep{romani_90,torres_16} or breakout of fields from the proto-neutron star convection zone. Regardless of these uncertainties, it is remarkable, though, that the weak-field model actually develops a strong surface dipole field by the end of the simulation than the strong-field model. As far as the proto-neutron star surface field is concerned, memory of the initial conditions appears to be erased within a few hundred milliseconds after collapse. This casts some doubt on the viability of the fossil-field scenario for magnetar formation. Whether amplification and reconnection processes in the supernova core indeed make the neutron star magnetic field independent of the initial field in the long term requires further investigation, however, both because of potential numerical uncertainties (resolution, treatment of the induction equation, etc.) and because the initial field in our simulation was still prescribed by hand. Aside from the question of neutron star surface fields, it is also difficult to square the high explosion energy of the strong-field model with the characteristics of supernova remnants around magnetars, which point to have normal or modest explosion energies \citep{vink_06}.  

Nonetheless, our simulations underscore that it is important to consider the role of magnetic fields in supernova explosions -- beyond the more spectacular scenario of magnetorotational supernovae -- in order
to more fully understand the explosion dynamics and the progenitor-remnant connection. As with hydrodynamic simulations, a more integrated view of magnetic field effects from the progenitor stage to the remnant phase is required for this purpose.

\section*{Acknowledgements}

BM acknowledges support by ARC Future Fellowship FT160100035. This work is based on simulations performed within computer time allocations from Astronomy Australia Limited's ASTAC scheme, the National Computational Merit Allocation Scheme (NCMAS), and an Australasian Leadership Computing Grant on the NCI NF supercomputer Gadi. This research was supported by resources provided by the Pawsey Supercomputing Centre, with funding from the Australian Government and the Government of Western Australia. This work has received funding from the European Research Council (ERC) under the European Union’s Horizon 2020 research and innovation programme (Grant agreement No.\ 945806).

%%%%%%%%%%%%%%%%%%%%%%%%%%%%%%%%%%%%%%%%%%%%%%%%%%
\section*{Data Availability}
The data underlying this article will be shared on reasonable request to the  authors, subject to considerations of intellectual property law.



\bibliographystyle{mnras}
\bibliography{paper}

%%%%%%%%%%%%%%%%%%%%%%%%%%%%%%%%%%%%%%%%%%%%%%%%%%

% Don't change these lines
\bsp	% typesetting comment
\label{lastpage}
\end{document}